\documentclass[lettersize,journal]{IEEEtran}
\usepackage{amsmath,amsfonts}
\usepackage{algorithm}
\usepackage{array}
\usepackage[caption=false,font=normalsize,labelfont=sf,textfont=sf]{subfig}
\usepackage{textcomp}
\usepackage{stfloats}
\usepackage{url}
\usepackage{verbatim}
\usepackage{graphicx}
\usepackage{cite}

\usepackage{algpseudocode}
\usepackage{booktabs}
\usepackage{threeparttable} 
\usepackage{hyperref}
\usepackage{orcidlink}
\usepackage{makecell}

\begin{document}

\title{A 1024-Channel 0.8V 23.9-nW/Channel Event-based Compute In-memory Neural Spike Detector}

\author{Ye Ke \orcidlink{0009-0002-9809-1192}, \IEEEmembership{Graduate Student Member, IEEE}, 
Zhengnan Fu \orcidlink{0009-0009-1235-8521}, \IEEEmembership{Graduate Student Member, IEEE}, 
Junyi Yang \orcidlink{0000-0002-5867-4943}, \IEEEmembership{Graduate Student Member, IEEE}, 
Hongyang Shang \orcidlink{0009-0007-6276-1947}, \IEEEmembership{Graduate Student Member, IEEE}, 
and Arindam Basu \orcidlink{0000-0003-1035-8770}, \IEEEmembership{Senior Member, IEEE}
\thanks{The authors are with the Department of Electrical Engineering, City University of Hong Kong, Hong Kong (email: arinbasu@cityu.edu.hk)}
\thanks{Manuscript received April 19, 2021; revised August 16, 2021.}}

\markboth{Journal of \LaTeX\ Class Files,~Vol.~14, No.~8, August~2021}%
{Shell \MakeLowercase{\textit{et al.}}: A Sample Article Using IEEEtran.cls for IEEE Journals}


\maketitle

\begin{abstract}
The increasing data rate has become a major issue confronting next-generation intracortical brain-machine interfaces (iBMIs). 
The scaling number of recording sites requires complex analog wiring and lead to huge digitization power consumption. Compressive event-based neural frontends have been used in high-density neural implants to support the simultaneous recording of more channels.
Event-based frontends (EBF) convert recorded signals into asynchronous digital events via delta modulation and can inherently achieve considerable compression. 
But EBFs are prone to false events that do not correspond to neural and may affect the output firing rate, which is the key feature for neural decoding.
Spike detection (SPD) is a key process in the iBMI pipeline to detect neural spikes and further reduce the data rate. 
However, conventional digital SPD suffers from the increasing buffer size and frequent memory access power, and conventional spike emphasizers are not compatible with EBFs.
In this work we introduced an event-based spike detection (Ev-SPD) algorithm for scalable compressive EBFs.
To implement the algorithm effectively, we proposed a novel low-power 10-T eDRAM-SRAM hybrid random-access memory (HRAM) in-memory computing (IMC) bitcell for event processing.
We fabricated the proposed 1024-channel IMC SPD macro in a 65nm process and tested the macro with both synthetic dataset and Neuropixel recordings. The proposed macro achieved a high spike detection accuracy of 96.06\% on a synthetic dataset and 95.08\% similarity and 0.05 firing pattern MAE on Neuropixel recordings.
Our event-based IMC SPD macro achieved a high per channel spike detection energy efficiency of 23.9 nW per channel and an area efficiency of 375 $\mu m^2$ per channel. 
Our work presented a SPD scheme compatible with compressive EBFs for high-density iBMIs, achieving ultra-low power consumption with an IMC architecture while maintaining considerable accuracy.
\end{abstract}


\section{Introduction}

\begin{figure}[!t]
    \centering
    \includegraphics[width=\columnwidth]{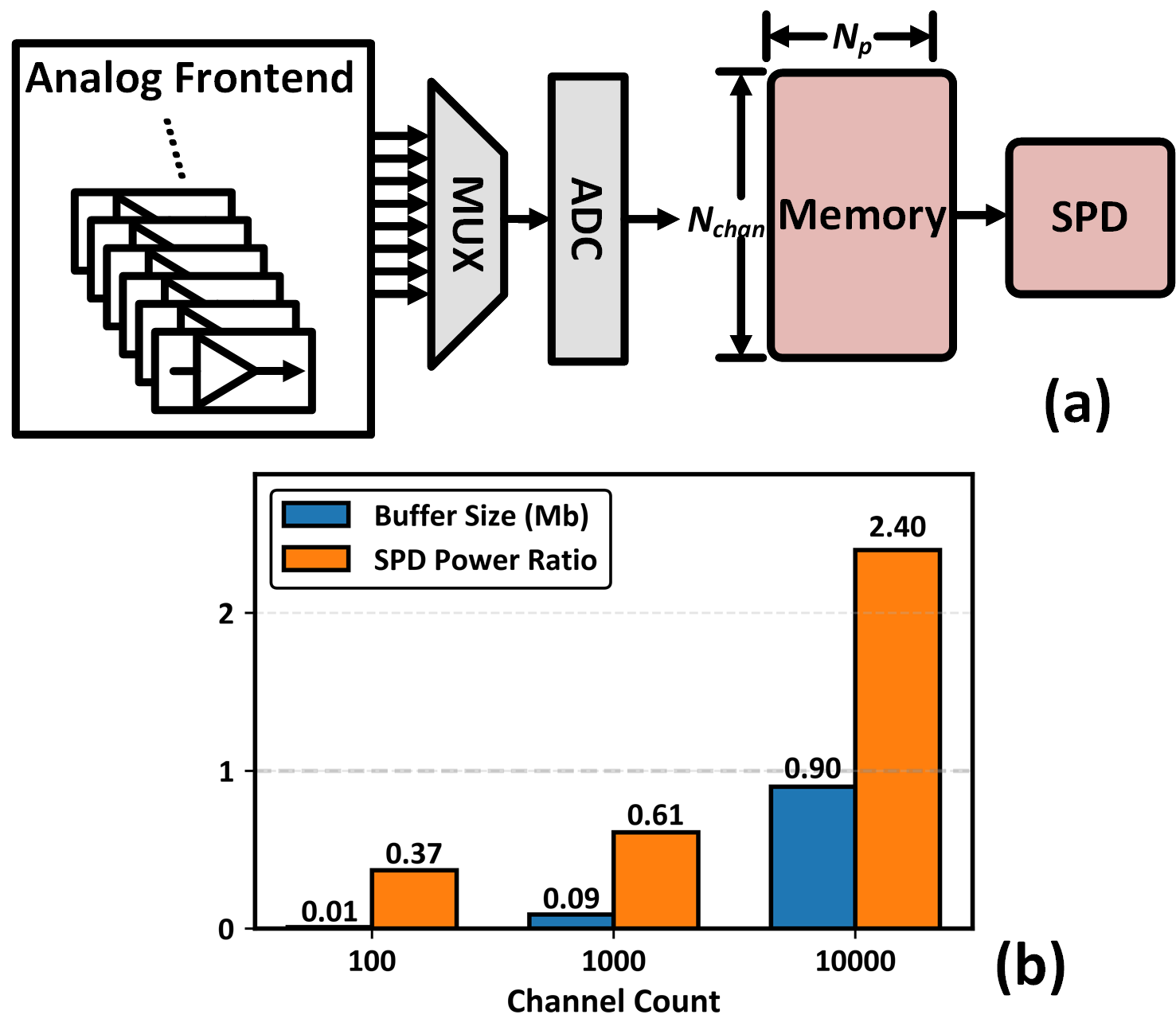}
    \caption{(a) Conventional Nyquist sampling iBMI spike detection pipeline. (b) The expanding buffer size for spike detection and power ratio between spike detection and frontend as the channel count scales in high-density iBMI.}
    \label{fig:Intro}
\end{figure}

\IEEEPARstart{I}{ntracortical} brain machine interfaces (iBMIs) have shown promising utilities in interfacing the human brain with external devices for prosthetic and cursor control\cite{lebedev_review}, speech decoding\cite{willettHighperformanceBraintotextCommunication2021}, and tactile sensations\cite{flesher_science} in the last few decades. Recent iBMI technologies are evolving towards higher channel counts to support the parallel recording of more neurons, which could bring about more robust and accurate neural decoding\cite{shaeri246mm$^2$MiniaturizedBrain2024}\cite{olearyBrainForestNeuromorphicMultiplierLess2025}\cite{chang15139mW200words2025}. 
But with the channel count of iBMIs scaling in a Moore-like trend\cite{kording_moore}, the exploding data rate has become a major challenge confronting next-generation iBMIs. 
First at the front-end, an increase in the number of recording sites would require more complex analog wiring. 
Consequently, digitizing, buffering and processing the enormous amount of raw neural data would lead to a huge power consumption, which may exceed the power budget of neural implants and damage brain tissue. 
Lastly the expanding data rate is beyond the capabilities of wireless transcutaneous neural implants, which could improve mobility and reduce the risk of infection\cite{mohanNeuromorphicCompressionBased2025}\cite{corradiNeuromorphicEventBasedNeural2015}.

To reduce the analog wiring and compress the neural signals at the front-end, a wired-or approach has been used in high-density front-ends\cite{jang1024Channel268nWPixel2023}\cite{akhoundi1521024Channel000029mm22025}
\cite{akhoundiScalable1024ChannelUltraLowPower2025}.
However, this lossy method could result in the loss of samples due to spatial collisions.
Inspired by dynamic vision sensors\cite{lichtsteiner128x128120DB2008}, event-based neural front-ends (EBFs) have been discussed in recent works to enable the simultaneous recording of over thousands of channels\cite{xu155EventBasedSpatially2025}\cite{corradiNeuromorphicEventBasedNeural2015}\cite{heEventBasedNeuralCompressive2024}\cite{heImplantableNeuromorphicSensing2022}.

Generally, the modulation thresholds are set such that noise alone does not trigger a threshold crossing, while action potentials typically cross this threshold. 
Compared to threshold crossing-based spike detectors, delta-modulating frontends can also preserve some information about the spike shape. 
However, due to the noisy nature of intracortical recordings, delta-modulation schemes are prone to falsely generating events that do not imply action potentials. 
These falsely generated events may affect the output firing rate, which is one of the key features used for neural decoding.

\begin{figure*}[t]
    \centering
    \includegraphics[width=0.95\textwidth]{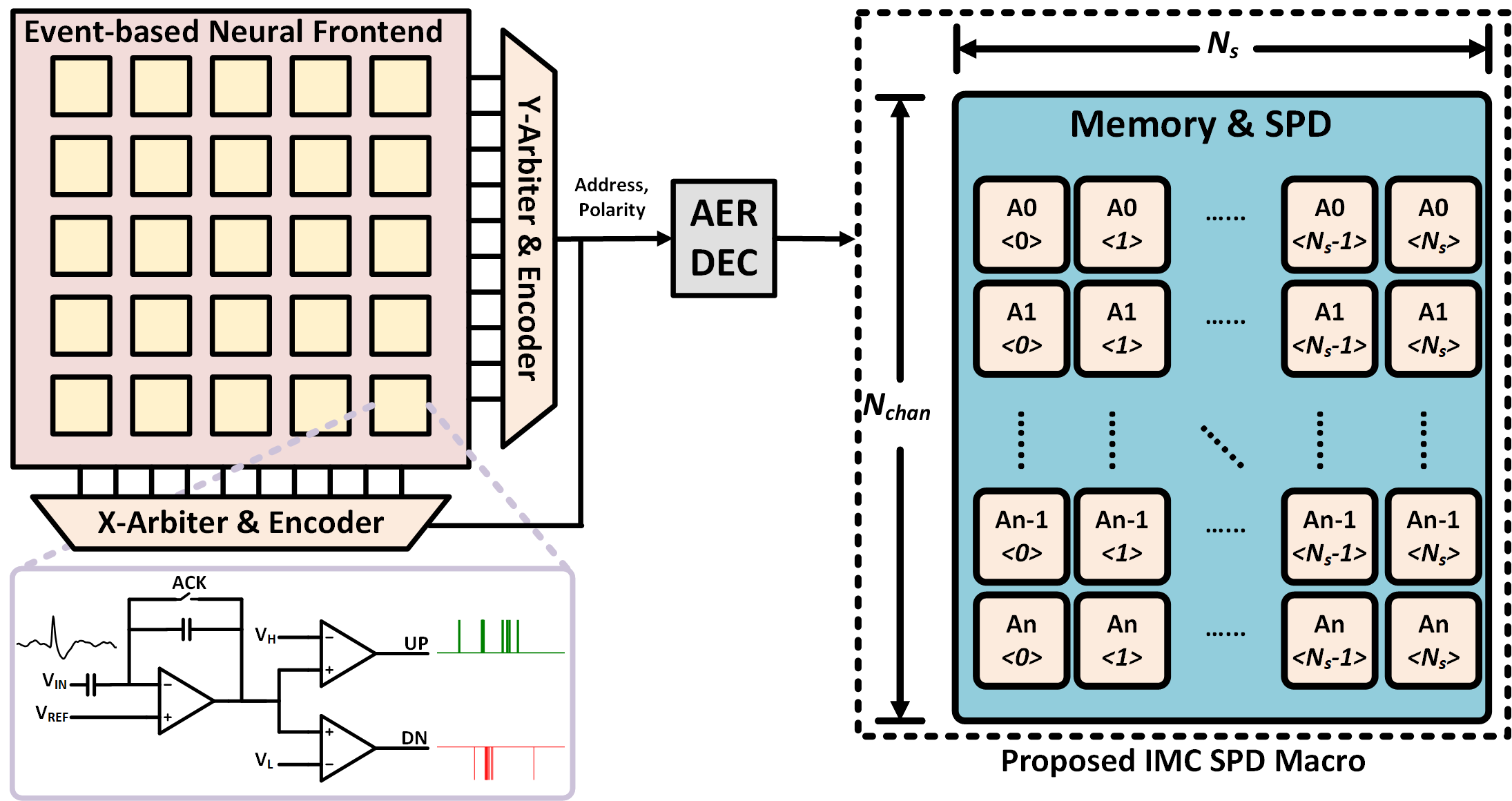}
    \caption{Proposed in-memory spike detection architecture for event-based neural frontends. The frontend delta modulator produces AER events when the delta of the input exceeds the threshold similar to the analog-to-spike conversion in \cite{cartiglia4096ChannelEventbased2024}. The AER event is decoded into the corresponding address of an $N_{chan}$ by $N_{s}$ memory array to implement the in-memory computations needed for spike detection.}
    \label{fig:Proposed architecture}
\end{figure*}

To distinguish neural spikes from noise in neural recordings, neural spike detection (SPD) is a key process in the iBMI signal processing pipeline (Fig. \ref{fig:Intro}(a)) and could further reduce the data rate profoundly. Conventional spike detection involves the buffering of digitized neural data samples and the real-time calculation and thresholding of signal emphasizers like nonlinear energy operators (NEO) \cite{mukhopadhyayNewInterpretationNonlinear1998}\cite{zhangAdaptiveSpikeDetection2021}. 
However, the buffer size needed for neural spike detection also scales with the channel count. Due to frequent memory access, the power dissipated by the SPD could grow more significant than amplifier power and become the dominant dissipation in the iBMI signal chain. 
As shown in Fig. \ref{fig:Intro} (b), when the channel count scales to 10k, the buffer size needed for SPD could reach $10  \text{ bits}\times 9 \text{ samples}\times 10k\text{ channels}= 0.9\text{ Mb}$ for the 10-bit, 9-sample digital SPD in \cite{biederman478Mm22015} and the power dissipation for the SPD alone could reach $\approx$24 mW (dominated by memory access estimated from \cite{horowitz11ComputingsEnergy2014}). Given that the power budget for a neural implant is $\approx$30 mW\cite{putzeysNeuropixelsDataAcquisitionSystem2019} and the neural amplifier for 10k channel requires $\approx$10 mW\cite{donghan045100ChannelNeuralRecording2013}, the memory access power in conventional Nyquist sampling architecture is unsustainable to support scaling to 10k recording channels.
Moreover, traditional spike detector implementations are not compatible with compressive EBFs that have better scalability at the frontend.

In an effort to reduce the memory access and the computation cost, in-memory computing (IMC) architecture have been widely discussed in recent studies \cite{heSRAMEDRAMBasedComputeinMemory2025}. 
\textcolor{black}{
However, previous in-memory computing arrays are designed for multiplication and accumulation operations.
To facilitate BMI signal processing with in-memory computing for compressive EBF,
we proposed a novel event-based in-memory computing array as depicted in Fig. \ref{fig:Proposed architecture}. 
We proposed a mixed-signal in-memory computing bitcell with an embedded DRAM that takes event-based input and implements in-bitcell computation of signal emphasizers, and an SRAM cell to store the results and calculate the moving sum.}
\textcolor{black}{Compared with the conventional digital implementation of NEO with normal SRAM sample buffers in \cite{biederman478Mm22015}, our IMC architecture reduces the memory array size for SPD by 9.5X, and achieves excellent per-channel SPD power efficiency.}
The proposed IMC macro enabled the effective processing of input event streams from EBF within the bitcell and supported the simultaneous event-based spike detection of 1024 neural channels at minimum power and area overhead.

The key contributions of this work include:
\begin{enumerate}
    \item We developed an accurate and efficient dual threshold (DT) spike detection algorithm compatible with compressive event-based neural frontends.
    \item To implement the proposed algorithm effectively, we proposed a novel 10-T eDRAM-SRAM hybrid random access memory (HRAM) IMC bitcell for processing events received from EBF.
    \item We presented an AER-compatible IMC array architecture for effective spike detection in high density iBMIs.
    \item We fabricated a 1024-channel IMC macro to validate the proposed event-based IMC SPD and evaluated the proposed event-based IMC SPD architecture on commonly used intracortical recording datasets.
\end{enumerate}

\textcolor{black}{
\section{Preliminaries}
\label{sec:preliminary}
\subsection{Event-based Frontend}
Figure \ref{fig:Proposed architecture} bottom presents a delta modulator circuit within a neural sensing pixel similar to the analog-to-spike conversion circuits in \cite{cartiglia4096ChannelEventbased2024,corradiNeuromorphicEventBasedNeural2015} (other architectures are also possible \cite{heEventBasedNeuralCompressive2024}).
As shown in the example waveforms, the input signal is converted into asynchronous digital events via delta modulation or its variants, inherently achieving considerable data compression at the front end, by leveraging the sparsity of action potentials.
Based on hardware measurement results from the latest event-based implantable frontends, EBFs have reported a low per-channel power of 0.8-1.38 $\mu$W/Ch, and compact per-channel area of 2304-3200 $\mu m^2$, which is superior to existing Nyquist frontends (with 5.9-42 $\mu$W/Ch power consumption in the same process\cite{cartiglia4096ChannelEventbased2024}\cite{chenNeuronInspired00032mm2138mW2024}) in terms of energy efficiency.
On the system level, the data rate reduction from EBF further reduces the energy consumption of communication \cite{mohanArchitecturalExplorationNeuromorphic2023}.
}

\textcolor{black}{
Asynchronous events are typically transmitted to the next stage through a communication protocol known as address-event representation (AER), which is commonly employed in the readout of dynamic vision sensors to effectively transmit sparse events with minimum wires \cite{gallegoEventBasedVisionSurvey2022} \cite{lichtsteiner128x128120DB2008}. 
The spike generated at a specific pixel is represented as a tuple of (address, polarity) via the X and Y address encoders, as shown in Fig. \ref{fig:Proposed architecture}.
The event arbiter can resolve the spatial collisions of EBF events at a typical delay of a few ns\cite{aungmyatthulinnAdaptivePriorityToggle2011}, which is minor compared to the typical spike durations. Effects of such collisions are analyzed in detail in \cite{mohanNeuromorphicCompressionBased2025} and show little degradation of RMSE after reconstruction. 
The 2D address encoding and compressed data rate from EBF significantly solves the routing congestion and reduces the encoding overhead compared with conventional Nyquist sampling. However, architectures for efficient on-chip processing of these events to perform SPD are still missing.
}
\textcolor{black}{
\subsection{eDRAM based IMC}
eDRAM-based bitcells for IMC have become popular due to their excellent energy and area efficiency\cite{heSRAMEDRAMBasedComputeinMemory2025}. The area efficiency generally stems from storing multi-bit values on a capacitor\cite{chen15365nm3T2021} compared to using multiple SRAM cells. While earlier voltage programming approaches\cite{chen15365nm3T2021} suffered from considerable variability, new current programming approaches \cite{song4bitCalibrationFreeComputingInMemory2024} have managed to overcome this using write feedback methods using a self-biased current mirror. However, these designs only target matrix vector multiply operations for neural networks and are not compatible with event-based circuits. Moreover, their retention time is typically few hundred $\mu$s even with added metal capacitors \cite{yuLogicCompatibleEDRAMComputeInMemory2021} making them unsuited for directly processing biomedical signals with longer time constants.}

\textcolor{black}{A hybrid eDRAM-SRAM approach was used in \cite{zhang9151220TOPS97613012023} to filter events from a dynamic vision sensor. This bitcell accumulated events on the eDRAM in the beginning while the latch was disabled by a transmission gate (TG) switch in series with the feedback inverter of the SRAM latch. This causes two issues: first, the sensing inverter dissipates a large amount of crowbar current since its input is an analog voltage on the eDRAM, and second, the junction leakage from the TG reduces the retention time making it unsuited for sparse events from biomedical sensors.}

\section{Computing-in-memory Event-based Spike Detection}
\label{sec:cim-spd}
\begin{figure}[t]
    \centering
    \includegraphics[width=\columnwidth]{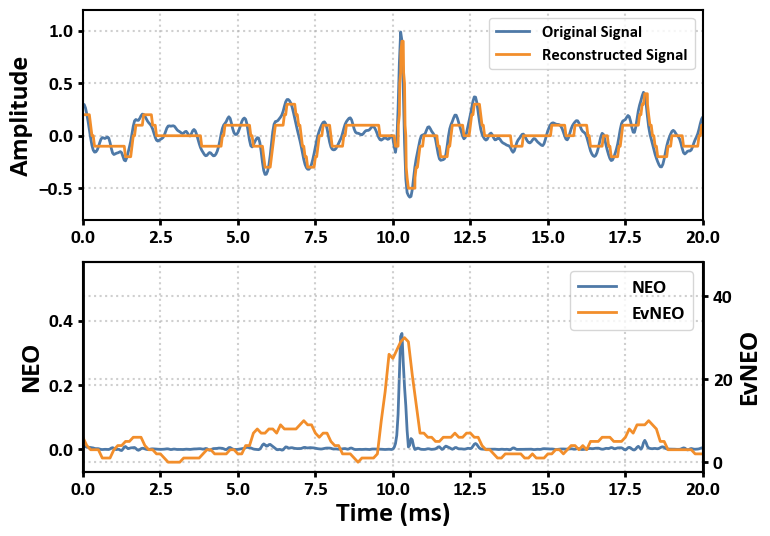}
    \caption{Top: The original signal from the dataset and the reconstructed signal from AER events. Bottom: standard NEO on original signal and proposed EvNEO on reconstructed signal.}
    \label{fig:EVNEO}
\end{figure}
In this section, we describe the proposed event-based spike detection algorithm (Ev-SPD) for compressive EBF without recovering a stair-step reconstruction. 
We compare the algorithm with conventional NEO to demonstrate excellent performance and discuss the design space explorations. 
Subsequently we present the eDRAM-SRAM hybrid in-memory computing bitcell to effectively implement the operations in the proposed algorithm. 
Then we introduce an AER-compatible 1024-channel IMC SPD macro for multi-channel in-memory event-based processing.

\subsection{Event-based Spike Detection Algorithm}\label{algorithm_sec}
\begin{figure*}[t]
    \centering
    \includegraphics[width=0.95\textwidth]{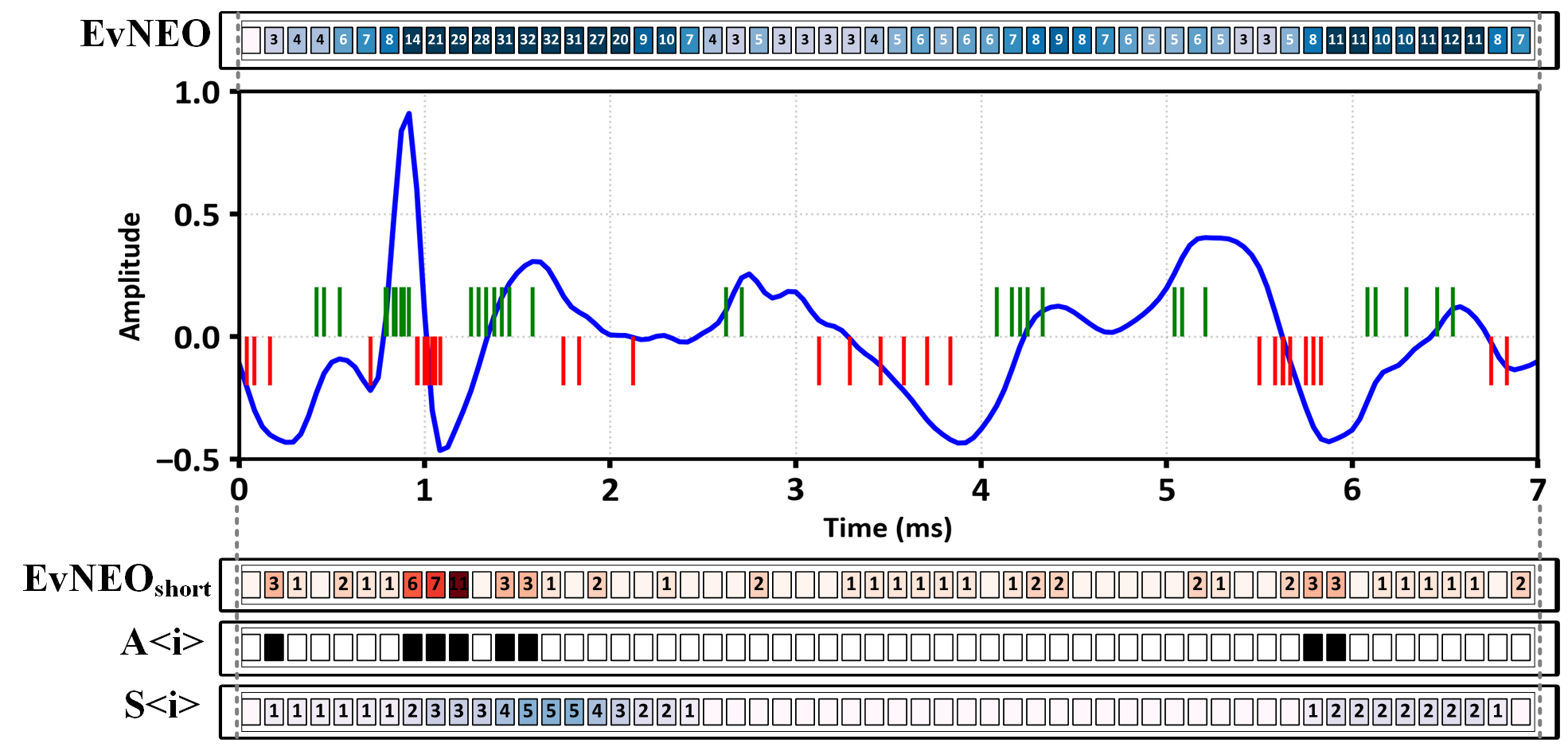}
    \caption{Top: the original signal, the delta modulated ON/OFF events from event-based neural frontend, and the $\operatorname{EvNEO}$ emphasizer values. Bottom: proposed event-based algorithm for IMC implementation.
    $\operatorname{EvNEO_{short}}$ denotes the event-based NEO approximation accumulated on eDRAM, which is then thresholded to get a binary sequence ${A}\langle{i}\rangle$. A sliding window moving sum is applied on ${A}\langle{i}\rangle$ to derive ${S}\langle{i}\rangle$ and then compared with a second threshold to give the spike detection result.}
    \label{fig:Algorithm}    
\end{figure*}

\begin{figure}[t]
    \centering
    \includegraphics[width=\columnwidth]{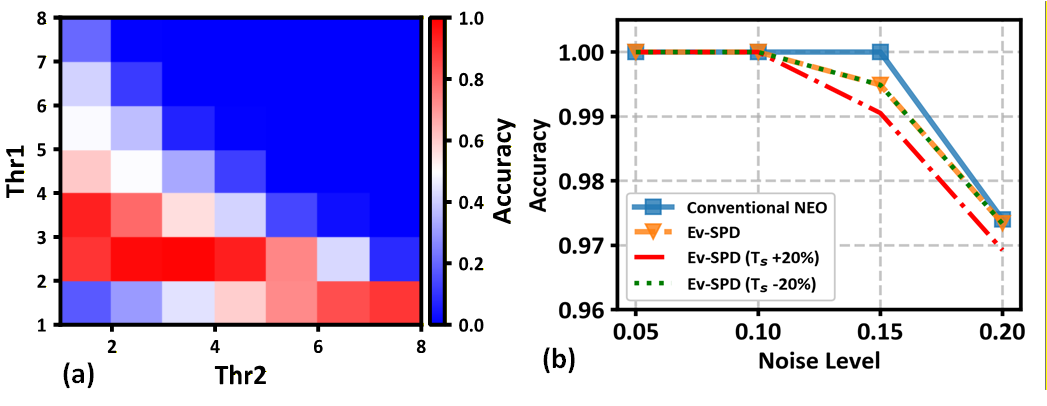}
    \caption{(a) Average spike detection accuracy of the simulated Ev-SPD algorithm on the synthetic dataset at different thresholds. \textcolor{black}{$Thr_1$ denotes the threshold to compare the accumulated voltage due to events in a short time window ($T_s$) to get the binary value ${A}\langle{i}\rangle$ (see Eq. \ref{eq:A_i}). $Thr_2$ denotes the threshold with which the moving sum $S\langle i \rangle$ is compared to derive the detection result.} (b) Comparison between the proposed event-based algorithm and the traditional nonlinear energy operator spike detection.}
    \label{fig:SimulationResults}
\end{figure}

\subsubsection{Ev-SPD Algorithm}
Previous neural spike detectors use spike emphasizers like nonlinear energy operator (NEO)\cite{mukhopadhyayNewInterpretationNonlinear1998} to enhance the SNR of action potentials and then threshold to detect the neural spikes. Previous work using EBF has shown that the spike information is retained in the generated events by reconstructing the neural spike\cite{mohanNeuromorphicCompressionBased2025} (e.g. Fig.\ref{fig:EVNEO} (top) shows a stair step reconstruction), which requires heavy computation and storage.
To perform SPD directly on delta-modulated events without recovering a stair-step reconstruction, we adopted a variant of NEO termed energy detection with low-pass filter (ED-LPF) \cite{yao0740NW2016} described in equation \ref{eq:ED-LPF}.
\begin{equation}
\label{eq:ED-LPF}
\operatorname{ED-LPF}\left(V_{in}(t)\right)=h(t)*\left(dV_{in}/dt\right)^2
\end{equation}
where $h(t)$ denotes the impulse response of a low-pass filter. 
Fig.\ref{fig:EVNEO} shows the proposed event-based approximation termed EvNEO described by:
\begin{equation}
\label{eq:NEO'}
\operatorname{EvNEO}\left(V_{in}(t)\right)=\int_{t-T_{spk}}^{t}\left|d \widehat{V}_{in}/dt\right|
\end{equation}
where $\widehat{V}_{in}$ denotes the stair-step reconstruction recovered from modulated events, ${T}_{spk}$ = 1 ms is the average spike duration, the integral is used to replace the low-pass filter and the squaring operation is replaced by the absolute value.

The benefit of this approximation is that $\widehat{V}_{in}/dt$ is directly available from ON/OFF events generated by the EBF and the absolute value operation is conveniently performed by ignoring event polarity. 
Intuitively, an embedded DRAM cell can perform this accumulation efficiently, but eDRAM lacks the dynamic range for multiple event inputs and loses information over long time ($\approx$ ms) due to leakage.
Hence, we split the duration ${T}_{spk}$ into $N_{s}$ non-overlapping bins of duration ${T}_{s}=\frac{T_{spk}}{N_s}$ each, and use separate eDRAMs to accumulate events for every ${T}_{s}$ bin to get $\operatorname{EvNEO_{short}}$. To avoid information leakage in the eDRAM over long periods and to provide noise robustness, the accumulated voltage in the capacitor can be converted to a binary value (${A}\langle{i}\rangle$) by comparing with a threshold \textcolor{black}{($Thr_1$)} and then stored in an SRAM as given by: 
\begin{equation}
\label{eq:A_i}
{A}\langle{i}\rangle = \begin{cases} 
1 & \text{if } \operatorname{EvNEO_{short}}\langle{i}\rangle > Thr_{1} \\
0 & \text{otherwise}
\end{cases}
\end{equation}
Figure \ref{fig:Algorithm} shows an example of a neural signal, ON and OFF spikes generated by a delta modulating EBF, the approximation of spike detector by $\operatorname{EvNEO}$ as described in Eq. \ref{eq:NEO'}, its shorter version $\operatorname{EvNEO_{short}}$ and the threshold values ${A}\langle{i}\rangle$. A good correlation is observed between large values of $\operatorname{EvNEO_{short}}$ and ${A}\langle{i}\rangle$.

To match the $\operatorname{EvNEO}$ values, the ${A}\langle{i}\rangle$'s from $N_s$ successive time bins are summed to form ${S}\langle{i}\rangle$ as given by: 
\begin{equation}
\label{eq:S_i}
S\langle i \rangle = \sum_{k=i-N_{s}+1}^{i} A\langle k \rangle
\end{equation}
The moving sum ${S}\langle{i}\rangle$ is compared with a second threshold \textcolor{black}{$Thr_2$} to further mitigate false positives and provide the spike detection result for time bin $ i$. The ${S}\langle{i}\rangle$ values can be seen to smoothly track variations in $\operatorname{EvNEO}$.

\subsubsection{Algorithm Simulation}
\textcolor{black}{
Typically, a sliding window length of $T_{spk} = 1$ ms is sufficient to provide discriminative information about an action potential. However, the algorithm performance is not very sensitive to this choice and it works well (only 0.4\% change in accuracy) even when $T_{spk}$ changes by $\pm$20\% as shown in Fig. \ref{fig:SimulationResults}(b). On the other hand,}
the choice of $N_{s}$ (and hence $T_s$) is important to optimize the performance of the proposed EvSPD. 
When ${T_{s}}$ is too long, the leakage due to eDRAM loses temporal information over time, and when ${T_{s}}$ is too short, the value of $\operatorname{EvNEO_{short}}$ is too small to distinguish between noise and a valid neural spike, and the memory size may need to be expanded as well.
\begin{figure*}[t]
    \centering
    \includegraphics[width=\textwidth]{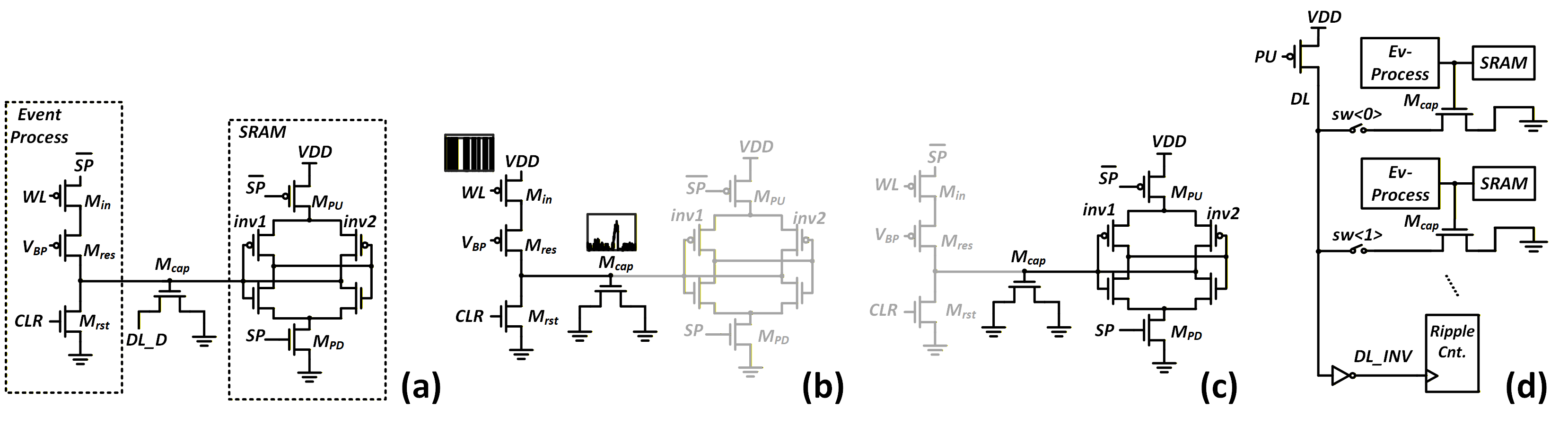}
    \caption{(a) Proposed 10-T IMC SPD bitcell. (b) Equivalent circuit of the bitcell working in accumulation phase, with an example of event pulse input and the voltage jump on eDRAM MOSCAP. (c) The equivalent circuit of the bitcell in the thresholding phase is as follows: the accumulated voltage is thresholded and stored in an SRAM latch. (d) The HRAM readout circuit, the channel-shared detection line is repeatedly precharged and sequentially connected to the MOSCAPs gated-controlled the SRAM.}
    \label{fig:Bitcell}  
\end{figure*}
\begin{figure*}[b]
    \centering
    \includegraphics[width=\textwidth]{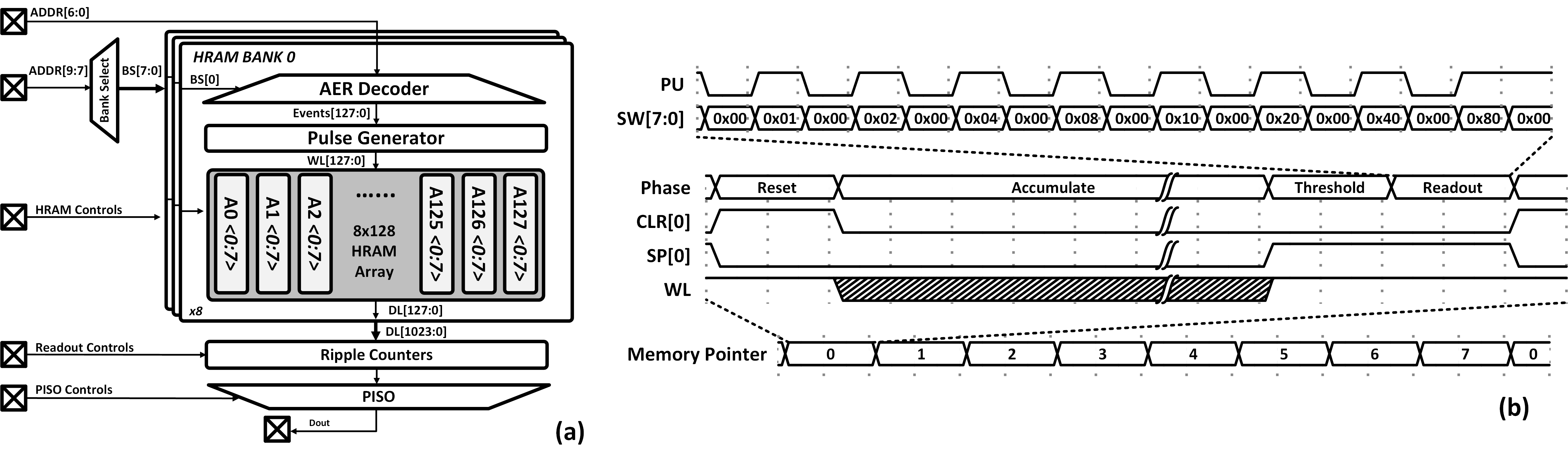}
    \caption{(a) System overview of proposed 1024-channel IMC SPD macro. The top three bits of the AER address are used as the bank select signal to enable one of the eight memory banks with each bank processing 128 neural channels. The HRAM values are readout to ripple counters through the detection lines and transmitted off-chip through a parallel-in-serial-out module.(b) Timing diagram of proposed HRAM bitcells. \textcolor{black}{A memory pointer selects the bitcell to update at the current detection period. Each detection period consists of sequential reset, accumulation, threshold, and readout phases. The bitcells are arranged as a circular buffer and the memory pointer wraps back to $0$.}}
    \label{fig:SystsemOverview}  
\end{figure*}
To perform a design space exploration, we applied the proposed Ev-SPD on a synthetic dataset \cite{quirogaUnsupervisedSpikeDetection2004} consisting of 4 recordings at 4 noise levels resulting in a total of 16 files to observe that a period of $T_{s}$ = 125 $\mu$s is sufficient to approximate $\operatorname{EvNEO_{short}}$ over a time bin, and a memory size of $N_{s}$ = 8 is selected to cover a spike duration of 1 ms in a sliding window manner. \textcolor{black}{In this case, the delta-modulation threshold is set at 0.1 time peak amplitude of spike to ensure good signal recovery and detection accuracy following \cite{mohanNeuromorphicCompressionBased2025}.} Furthermore, we investigate the sensitivity of spike detection accuracy to the choice of the two thresholds, $Thr_1$ and $Thr_2$. Figure \ref{fig:SimulationResults} (a) presents the SPD accuracy averaged on the dataset \textcolor{black}{(described in Sec. \ref{sec:results})} at different threshold settings to show robust detection accuracy of the proposed SPD across a wide range of threshold values. 
Further, in Fig. \ref{fig:SimulationResults} (b), we compared the proposed event-based algorithm with a conventional NEO-based SPD \textcolor{black}{with the empirical threshold setting in previous hardware implementations\cite{zhangCalibrationFreeHardwareEfficientNeural2023}} to show similar spike detection performance across different noise levels. The results confirm that the event-based approximation proposed here is a good approximation to the original one. \textcolor{black}{
Lastly, to verify the robustness of the proposed method to variations in the front-end modulation factor, we conducted a sensitivity analysis by changing the modulation factor by $\pm$20\%. The accuracy decreased by at most 0.3\%, demonstrating the proposed algorithm’s robustness to variations in the modulation factor.
 In the appendix, we present the results of applying our algorithm on another synthetic dataset\cite{buccinoMEArecFastCustomizable2021} with varying SNR to show the generality of our method.}

\begin{figure}[t]
    \centering
    \includegraphics[width=\columnwidth]{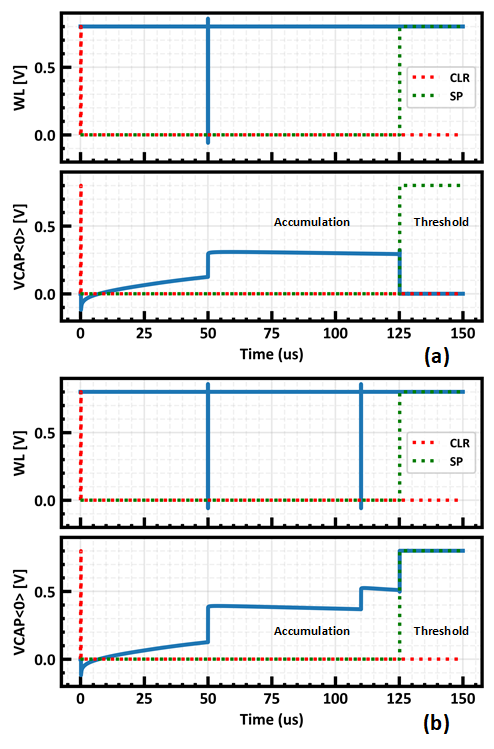}
    \caption{\textcolor{black}{(a) Example input, control signals, and capacitor voltage for a negative detection during the accumulation phase and thresholding phase. (b) Example input, control signals, and capacitor voltage for a positive detection during the accumulation phase and thresholding phase.} }
    \label{fig:Transient}  
\end{figure}

\begin{figure}[!b]
    \centering
    \includegraphics[width=\columnwidth]{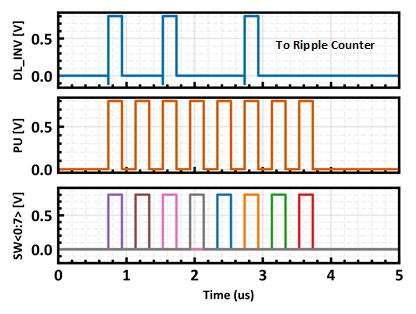}
    \caption{\textcolor{black}{Transient readout waveforms of pre-charge control signal PU, detection line switches controls SW and the inverted detection line as input to the ripple counter (see Fig. \ref{fig:Bitcell}(d)).}}
    \label{fig:readout}
\end{figure}

\subsection{eDRAM-SRAM Hybrid IMC Bitcell}
As stated in the previous section, the event accumulation operation can be implemented by an embedded-DRAM while an SRAM can be used to threshold and store the results for moving sum. 

We proposed a novel 10-T eDRAM-SRAM hybrid IMC bitcell as shown in Fig. \ref{fig:Bitcell}. This bitcell could effectively implement this SPD with $N_s = 8$ bitcells each channel.  
The left part of the bitcell is a 3T1C eDRAM for event processing, consisting of: an input transistor $M_{in}$, a pull-up variable resistor $M_{res}$, a MOS capacitor $M_{cap}$, and a pull-down reset transistor $M_{rst}$. 
\textcolor{black}{
At the beginning of each accumulation phase, the DRAM is reset via the transistor $M_{rst}$.
Then the events from EBFs are input through $M_{in}$ to enable event-based accumulation. Each input event is converted into a voltage jump on the eDRAM capacitor $M_{cap}$, with the magnitude controlled by the pull-up PMOS $M_{res}$. The accumulated voltage on the MOSCAP represents the $\operatorname{EvNEO_{short}}$ presented in Section \ref{algorithm_sec}. 
The right side of the bitcell schematic is a 6T SRAM latch, consisting of two inverters and two power-gating transistors. 
After the eDRAM finishes accumulating in the current time bin, the SRAM is enabled to latch the result into the binary value $A\langle i \rangle$ for time bin $i$. The threshold of $inv1$ acts as the binarization threshold $Thr_1$.
}
\textcolor{black}{Compared to previous hybrid bitcells\cite{zhang9151220TOPS97613012023} for DVS applications described in Section \ref{sec:preliminary}, our power-gating design could significantly reduce the crowbar current induced power consumption of the SRAM inverters (by $>10^4$X) and largely reduced the leakage during eDRAM accumulations by removing the TG switch.}
In addition, the drain of the MOS capacitor can be configured to either ground or connect to a channel detection line, allowing the transistor to be alternatively used as a readout transistor gate controlled by the SRAM.

\subsection{Operation Phases}
Corresponding to the accumulation, thresholding and moving sum operations in the proposed EvNEO algorithm, the HRAM IMC SPD bitcell has four sequential operation phases in a detection cycle: reset, accumulation, thresholding, and readout, as shown in the timing diagram Fig. \ref{fig:SystsemOverview} (b). \textcolor{black}{The eight bitcells for a neural channel are organized as a circular buffer to store eight previous ${A}\langle{i}\rangle$ values.}

\subsubsection{Reset}
At the beginning of a detection period, one of the eight bitcells is selected for update. \textcolor{black}{The index of the bitcell to be updated is indicated by a rotating memory pointer as shown in Fig. \ref{fig:SystsemOverview} (b) bottom.}
$M_{cap}$ of the corresponding bitcell is reset by the pull-down transistor.
\subsubsection{Accumulation}
After resetting, the SRAM latch of the selected bitcell is disabled by setting the power gating signal SP to 0. Both drain and source of $M_{cap}$ are grounded to use it as a MOSCAP as in Fig. \ref{fig:Bitcell} (b).
The events arriving in address event representation (AER) format are decoded by a decoder to trigger a pulse generator and produce a pulse on the input word line.
Each pulse input at the word line would result in a voltage jump on $M_{cap}$, determined by the configuration of $M_{cap}$ and $M_{res}$. 
The accumulated voltage of all input pulses over  ${T}_{s}$ approximates $\operatorname{EvNEO_{short}}$. 
The word line events missed in other phases are negligible since the time spent in the other phases is only 0.64\% of the accumulation phase.

\subsubsection{Thresholding}
After accumulation, the SRAM of the selected bitcell is enabled by setting SP to 1 to threshold and store the accumulated voltage as a binary value, as indicated in Fig. \ref{fig:Bitcell} (c). 
The inverter ${inv1}$ is designed to be strong to sense the voltage stored on $M_{cap}$.
The trip point of the SRAM is designed to apply a binary thresholding operation on $\operatorname{EvNEO_{short}}$. 
The accumulated voltage is then converted into a binary value indicating the presence of an action potential. The SRAM holds its value until the next accumulation cycle. 
\textcolor{black}{
Figure \ref{fig:Transient} shows the transient simulation waveforms for the accumulation and thresholding phases. Figure. \ref{fig:Transient}(a) presents an example of WL inputs and capacitor voltage for a negative detection. Fig. \ref{fig:Transient}(b) shows the corresponding waveforms for a positive detection.
}

\subsubsection{Readout}
After the selected bitcell is updated, multiple SRAM values are readout through a channel-shared detection line and a ripple counter for a digital summing operation as in Fig. \ref{fig:Bitcell} (d). 
The detection line is firstly pre-charged and then the drain of Mcap is connected to the detection line through a switch. 
The data stored in the SRAM determines whether the bitcell would discharge the detection line and produce a transition at the ripple counter clock. 
The same operation occurs for all eight bitcells in one channel and the ripple counter output is compared with a digital threshold to give the final detection result.
\textcolor{black}{
Figure \ref{fig:readout} presents the example transient readout waveforms of PU, SW and the inverted DL as input to the readout ripple counter.
}
\begin{figure}[!b]
    \centering
    \includegraphics[width=\columnwidth]{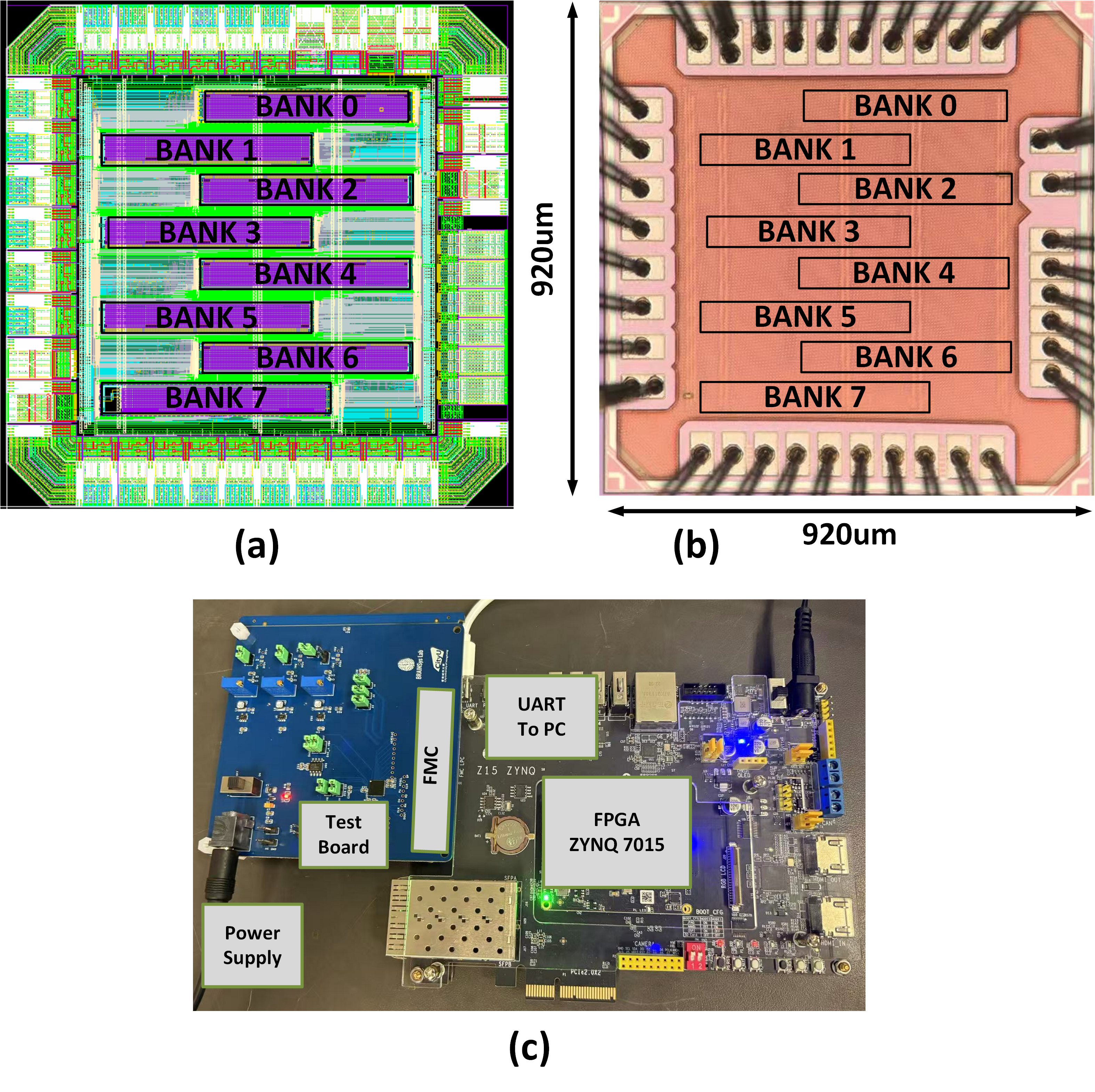}
    \caption{(a) Layout (b) Die micrograph of fabricated IMC SPD macro. (c) Chip testing setup: the delta-modulated datasets are interfaced to the test board through a Zynq 7015 FPGA. The FPGA reads out the HRAM values and sends the data to PC.}
    \label{fig:DieMicrograph}
\end{figure}

\begin{figure}[!b]
    \centering
    \includegraphics[width=\columnwidth]{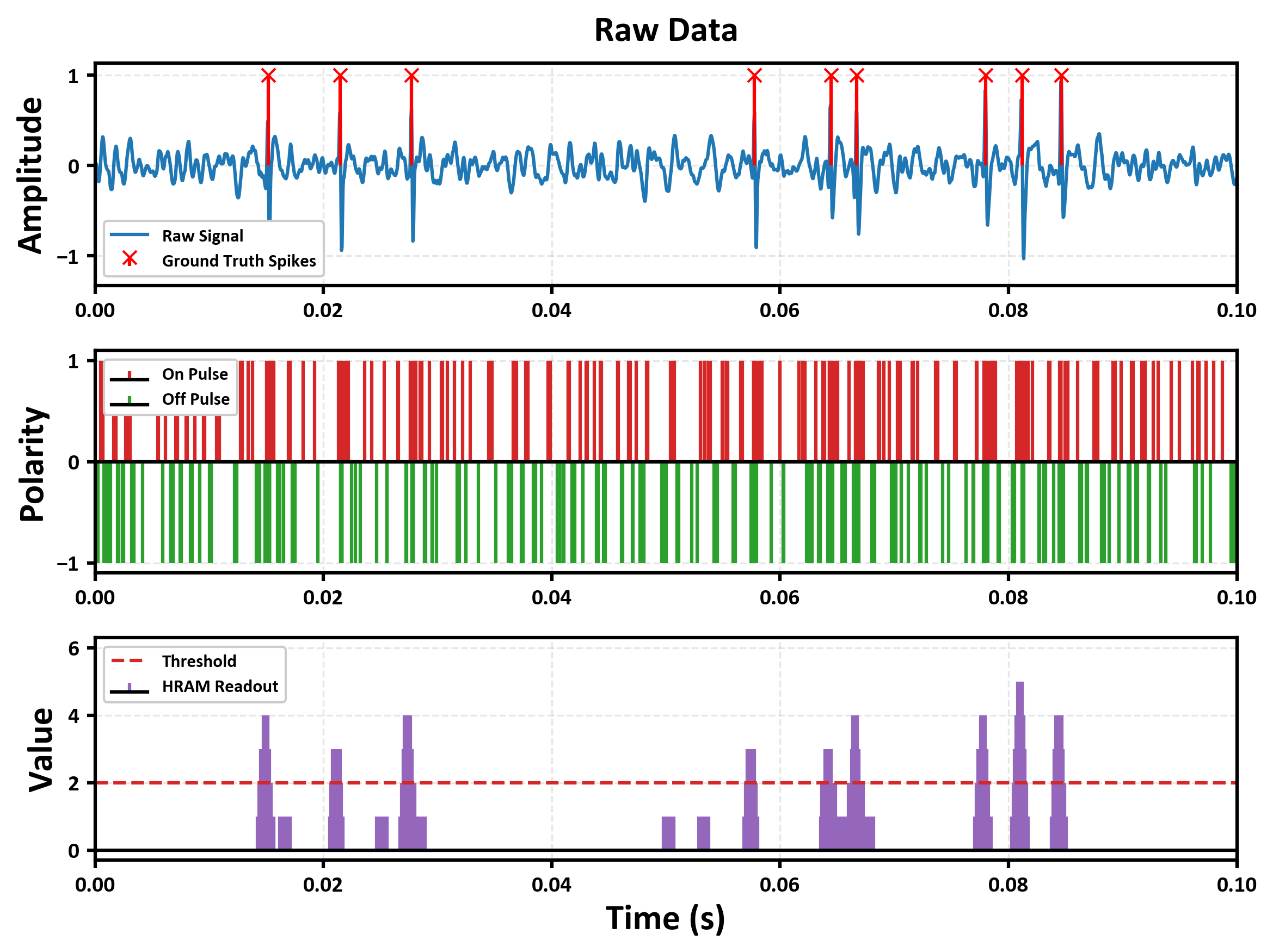}
    \caption{(a) Neural recordings and ground truth from the dataset. (b) Delta-modulated AER event streams. (c) Ripple counter readouts for each detection cycle from the chip.}
    \label{fig:Examplewaveform}
\end{figure} 

\subsection{System Overview}
Figure \ref{fig:SystsemOverview} (a) introduces the top-level architecture of the proposed 1024-channel in-memory computing neural spike detector, featuring an AER-compatible interface and an 8x1024 IMC array divided into 8 banks for effective in-memory event-based SPD. 
\textcolor{black}{
In the implementation, we layout the 8x128 memory bank as a hard macro and integrate eight memory banks with surrounding logic in the digital flow.
}
The IC supports up to 10-bit AER address inputs with the top 3 bits as a bank selection signal. 
The input event channel address is decoded, triggering a pulse generation circuit that activates the word line of the corresponding memory column for event-based processing. 
Each column consists of eight HRAM bitcells configured logically as a circular buffer to detect spikes on one neural channel using a sliding window of $T_{spk}$ = 1 ms. 
A memory pointer will decide which of the 8 rows to be written at a certain detection period. 
The binary values stored in the HRAMs are summed by the ripple counter on each channel and then transmitted off-chip through a parallel-in-serial-out (PISO) module.

\section{Measurement Results}
\label{sec:results}

\begin{algorithm}[!b]
\caption{Delta Modulation}\label{dm_algo}
\begin{algorithmic}[1]
\Require 
    \Statex $signal$: input signal vector $[x_1, x_2, \dots, x_N]$
    \Statex $threshold$: delta modulation threshold
\Ensure 
    \Statex $pulses$: sequence of pulses coded as [time, polarity]
    
\State Initialize $pulses \gets \emptyset$ 
\State $V_{reset} \gets 0$
\For{$i \gets 1$ \textbf{to} $\mathrm{length}(signal)$}
    \State $\Delta \gets signal[i] - V_{reset}$
    \If{$\Delta > threshold$}
        \State $n \gets \left\lfloor \Delta / threshold \right\rfloor$
        \State $t \gets \mathrm{linspace}(i,\ i + 1,\ n)$
        \State Append $(t, 1)$ to $pulses$  \Comment{Positive pulses}
        \State $V_{reset} \gets V_{reset} + n \cdot threshold$
    \ElsIf{$\Delta < -threshold$}
        \State $n \gets \left\lfloor \Delta / -threshold \right\rfloor$
        \State $t \gets \mathrm{linspace}(i,\ i + 1,\ n)$
        \State Append $(t, -1)$ to $pulses$  \Comment{Negative pulses}
        \State $V_{reset} \gets V_{reset} - n \cdot threshold$
    \EndIf
\EndFor
\State \Return $pulses$
\end{algorithmic}
\end{algorithm}

We fabricated the proposed 1024-channel IMC spike detector macro in 65-nm CMOS process. Figure \ref{fig:DieMicrograph} (a) and (b) shows the layout and die micrograph of the macro, occupying a core area of 0.384 mm\textsuperscript{2}. In this section we tested the SPD performance of proposed IMC macro on two datasets to validate the proposed IMC SPD.
\subsection{Experiment Setup}

To validate the proposed IMC macro, we tested the SPD performance on two datasets: 
1) The synthetic dataset first introduced in \cite{quirogaUnsupervisedSpikeDetection2004} is widely used for spike detection and spike sorting algorithm evaluations. 
This dataset is composed of recorded spike waveforms arriving in a Poisson process. The dataset provides four groups of signals sampled at 24 kHz: Easy1, Easy2, Difficult1 and Difficult2. 
Within each group, the standard deviation of Gaussian noise ranges from 0.05 to 0.2 at a step of 0.05 relative to the peak amplitude. 
2) The Neuropixel dataset provided neural signals from the visual cortex of a mouse recorded by Neuropixel probes\cite{steinmetzNeuropixels20Miniaturized2021}. This electrode array recorded 384 channels of neural data at 30 kHz.

Both datasets are band-pass filtered to remove the LFP and then converted into AER events using the delta modulation algorithm introduced in Algorithm \ref{dm_algo}.

\textcolor{black}{
For the synthetic dataset\cite{quirogaUnsupervisedSpikeDetection2004}, the delta-modulation threshold is set as described earlier in Sec. \ref{sec:cim-spd}.
For Neuropixel recordings with large spike amplitude variations and low SNR, we first used an absolute threshold spike detection to extract 1-ms spike waveforms on each channel. 
The delta modulation threshold is empirically set to 0.5 times the peak-to-peak amplitude of the spike waveform, thereby constraining the event rate to $\approx$ 2.5 kHz events per second per channel to achieve better compression and reduce the number of noise events on intracortical recordings. For both datasets, we have verified that the performance is not very sensitive to the choice of threshold with a maximum drop in accuracy of only $\approx0.3\%$ for $\pm20\%$ change in the threshold.
}

The AER events are quantized into timestamps and stored in the BRAM of the FPGA, and then they are fed into the chip to emulate an event-based neural frontend as shown in Fig. \ref{fig:DieMicrograph} (c). 
Meanwhile, the FPGA reads out the ripple counter values sequentially through the PISO module and then transmits them to the PC for SPD performance evaluation.
Figure \ref{fig:Examplewaveform} shows an example waveform from the synthetic dataset, the converted AER events, and the ripple counter readout from the chip.

\begin{figure}[!t]
    \centering
    \includegraphics[width=\columnwidth]{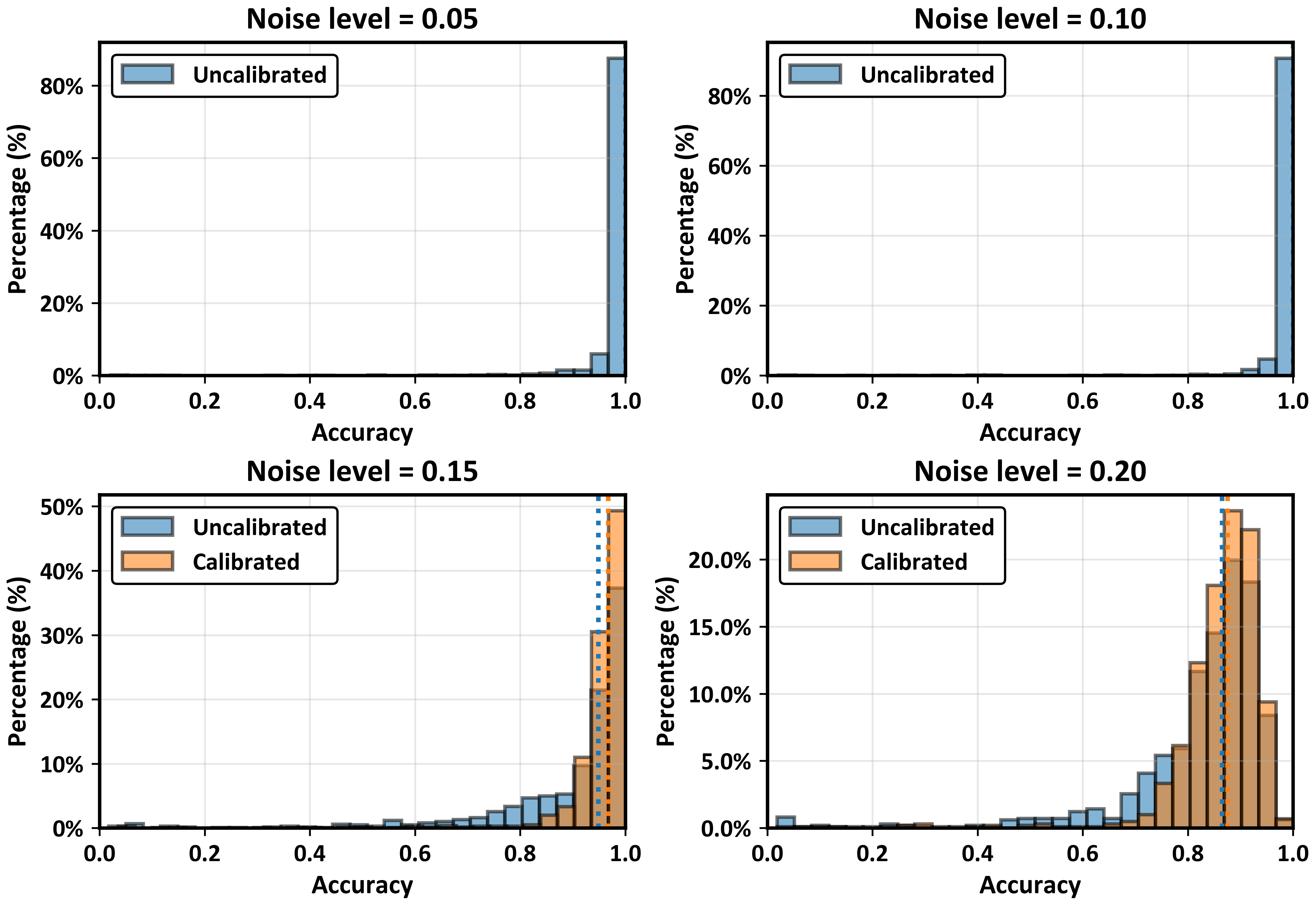}
    \caption{Measured 1024-channel spike detection accuracy on the synthetic dataset at different noise levels. Without calibration, the accuracy degrades at higher noise levels leading to 20.76\% of channels with accuracy less than 0.8. After calibration with 2 bits per channel, only 7.42\% remain in this category.}
    \label{fig:SyntheticMultichan}
\end{figure}
\begin{figure}[!t]
    \centering
    \includegraphics[width=\columnwidth]{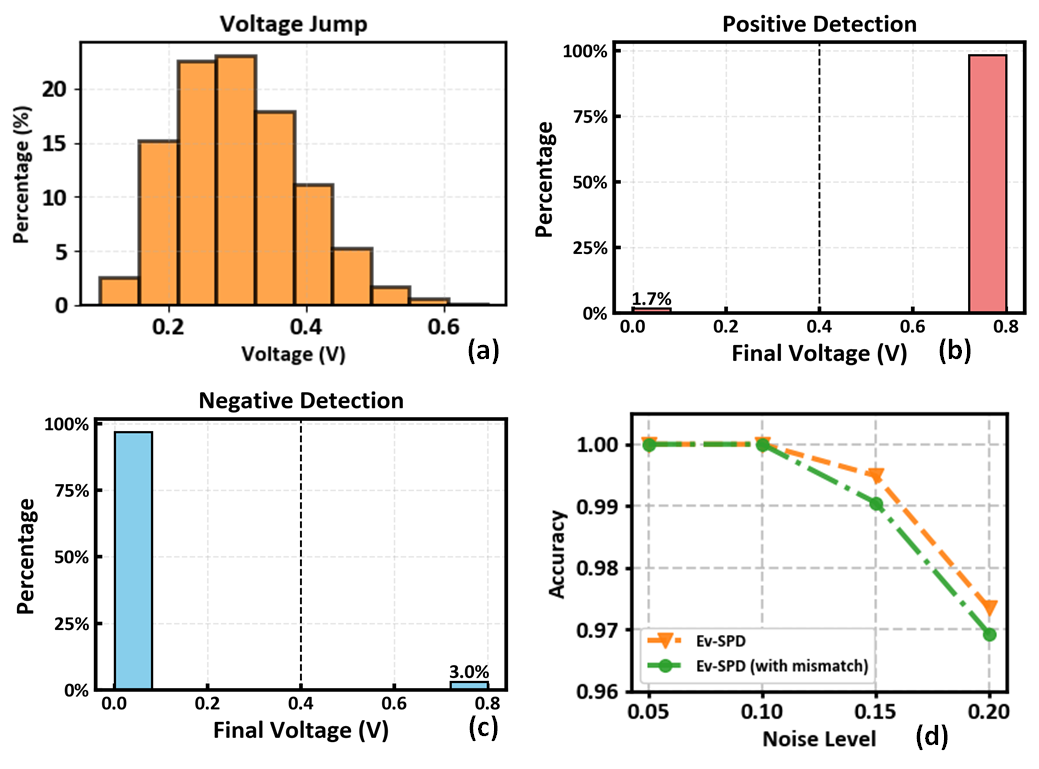}
    \caption{(a) 2000-run Monte Carlo simulation of eDRAM MOSCAP voltage jump for a unit pulse input.
    \textcolor{black}{
    (b) 2000-run Monte Carlo simulation of transient positive detection.
    (c) 2000-run Monte Carlo simulation of transient negative detection.
    (d) Comparison between the proposed event-based algorithm with and without Monte-Carlo mismatch.}
    }
    \label{fig:MCandSyntheticResult}    
\end{figure}
\begin{figure}[!t]
    \centering
    \includegraphics[width=\columnwidth]{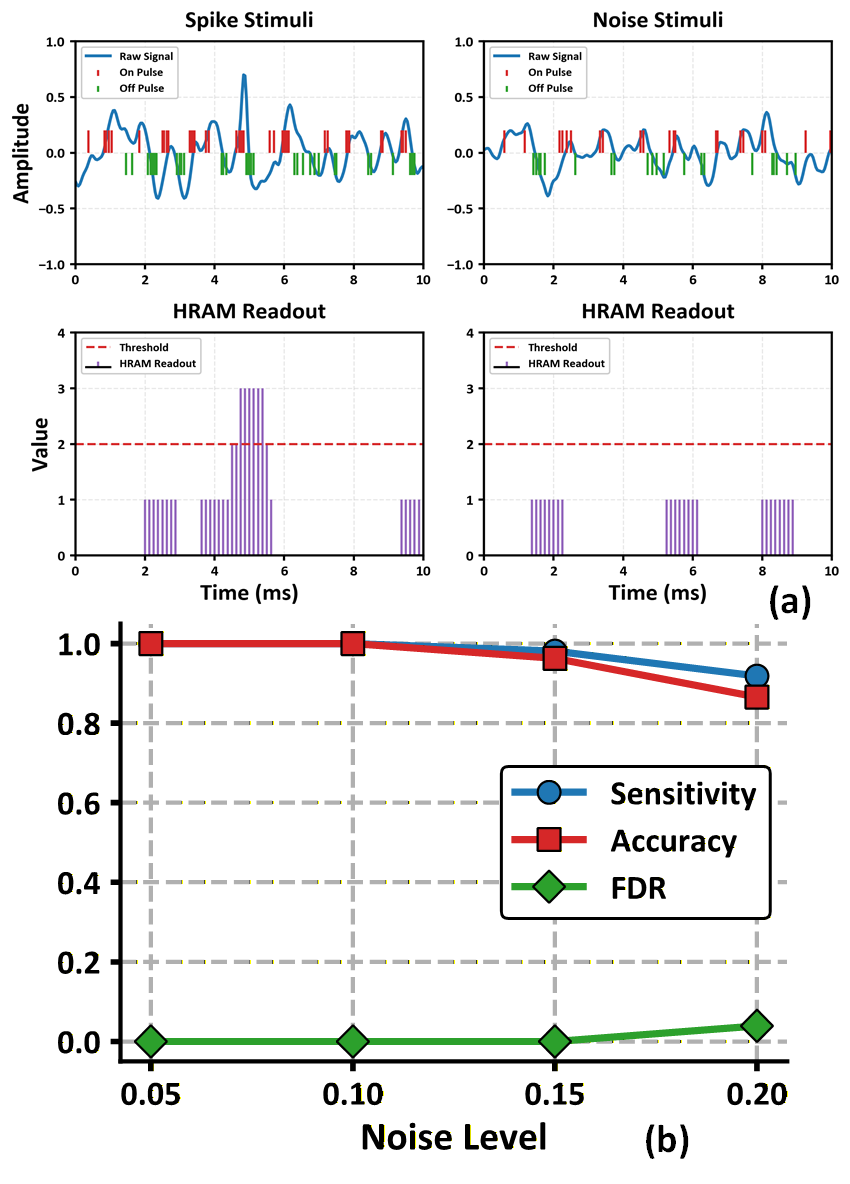}
    \caption{(a) Spike and noise stimuli for calibration and channel readouts. (b) Summary of measured spike detection metrics after calibration on the synthetic dataset.}
    \label{fig:calibration}
\end{figure}

\subsection{Spike Detection Performance}
\subsubsection{Synthetic Dataset}

Since the synthetic dataset is a single-channel dataset, we swept the single-channel recordings in the dataset across all fabricated IMC channels to estimate the mismatch across channels. 
We applied a digital threshold on the ripple counter readouts from each channel and added a refractory period of 1 ms. Then event time detected by the macro is compared with the ground truth spike timestamps provided in the dataset.

\begin{figure}[t]
    \centering
    \includegraphics[width=\columnwidth]{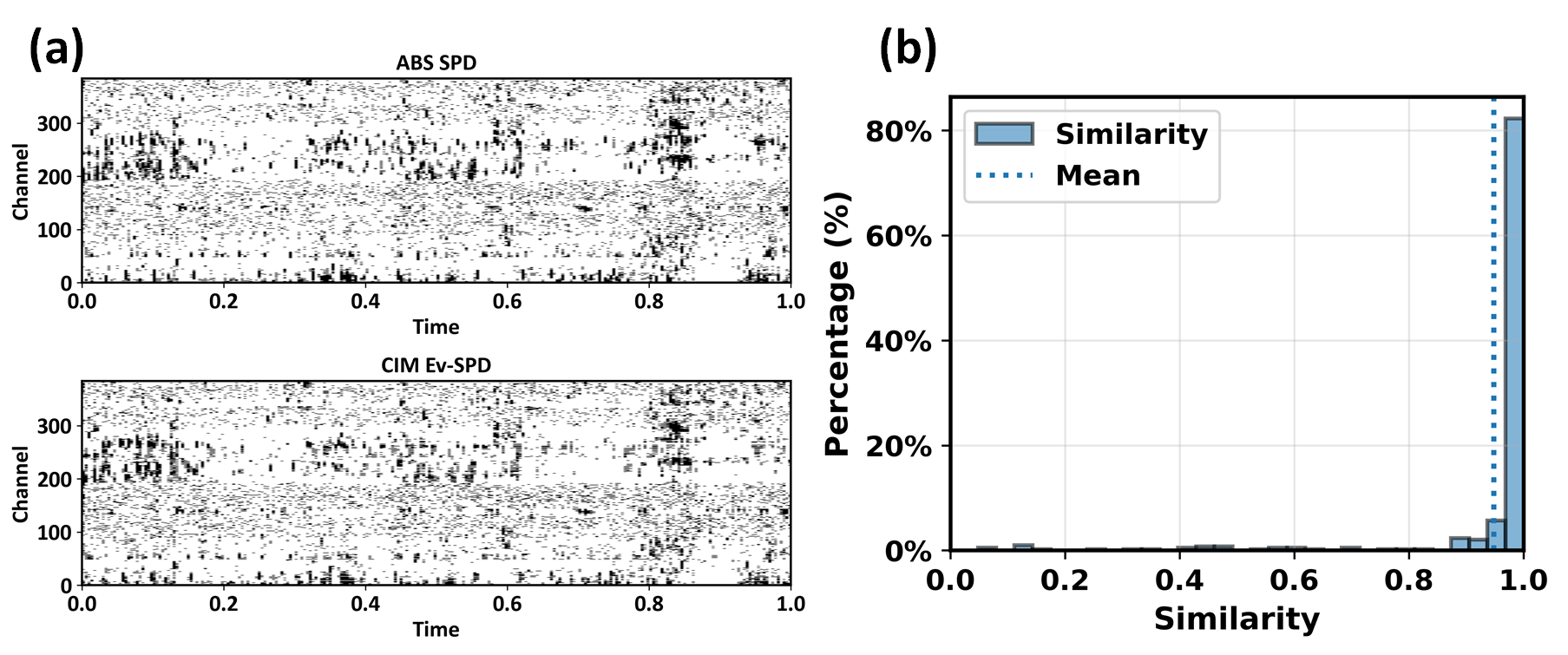}
    \caption{(a) Comparison of 384-channel Neuropixel recording firing rate patterns from absolute thresholding and IMC Ev-SPD. (b) 384-channel Neuropixel recording spike detection similarity compared with absolute thresholding SPD.}
    \label{fig:NP}
\end{figure}

\begin{table*}[!b]
\huge
\centering
\caption{Comparison with prior spike detectors}
\vspace{-20pt}
\resizebox{\textwidth}{!}{
\begin{threeparttable}
\label{tab:comparison}
\begin{tabular}{lcccccccc}
\toprule
 & \textbf{This Work} & 
 \textbf{ISSCC25\cite{akhoundi1521024Channel000029mm22025}} & \textbf{CICC24\cite{chenNeuronInspired00032mm2138mW2024}} & \textbf{VLSI24\cite{choi$Delta$BasedSpikeSorting2024}} & \textbf{TBCAS23\textsuperscript{b}\cite{zhangCalibrationFreeHardwareEfficientNeural2023}} & \textbf{TBCAS25\textsuperscript{b}\cite{huPowerandAreaEfficientChannelInterleavedNeural2025}} &
\textbf{JNE22\textsuperscript{b}\cite{valenciaVivoNeuralSpike2022}} &
\textbf{TBCAS20\cite{fiorelliChargeredistributionBasedQuadratic2020}}
 \\
\midrule
Technology (nm) & 65 & 40 & 40 & 65 & 65 & 65 & 180 & 180\\

Implementation & Mixed & Digital & Digital & Analog & Digital & Digital & Digital & Analog\\

Supply Voltage (V) & 0.8 & 0.72 & 1.2/0.6 & -- & 1.2 & 1.2 & 1.8 & 0.5\\
Detection Method & Event-based ED-LPF & VC and NEO& Integrate and fire & Delta threshold & NEO & NEO & NEO & MAE\\

Channel Count & \textbf{1024} & 1024 & 32 & 32 & 256 & 16 & 128 & --\\
Area per Channel ($\mu m^2$ /Ch) & 375&142\textsuperscript{a} & -- & -- & 751 & --  & $2\times10^{4}$&  $2.7\times10^{5}$\\

Power per Channel (nW/Ch) & \textbf{23.9} & 37.8\textsuperscript{a} & 480\textsuperscript{a}& 765\textsuperscript{a}& 38 & -- & $4.9\times10^{3}$ & 116\\

Neural Frontend & \makecell{\textbf{Delta modulation}} & Wired-or &\makecell{Delta modulation} & Delta-spikes &  \makecell{Nyquist sampling} & \makecell{Nyquist sampling} & \makecell{Nyquist sampling} & \makecell{Analog spikes}\\

AER Compatible & \textbf{Yes} & No &Yes & No & No & No & No & No\\
Detection Accuracy\textsuperscript{c} &  88--99\% & -- &-- & --& 91--98\% & 96--98\% & 92\% & -- \\
\bottomrule
\end{tabular}
\begin{tablenotes}
\item[a] {Spike detection power/area from system breakdown.}
\item[b] {Numbers obtained from ASIC simulation.}
\item[c] {Range of accuracy or mean accuracy measured on synthetic dataset}
\end{tablenotes}
\end{threeparttable}
}
\end{table*}

To quantitatively evaluate the spike detection algorithm in the synthetic dataset, we used three detection metrics: Sensitivity, False Detection Rate(FDR), and Accuracy calculated as below:
\begin{equation}
\text {Sensitivity}  =\frac{TP}{TP+FN}
\end{equation}
\begin{equation}
\text {FDR}  =\frac{FP}{TP+FP}
\end{equation}
\begin{equation}
\text {Accuracy}  =\frac{TP}{TP+FN+FP}
\end{equation}
In which True Positive (TP) denotes the number of spikes correctly identified by the algorithm;
False Positive (FP) denotes the number of non-spike inputs being falsely identified as spikes by the algorithm;
and False Negative (FN) denotes the number of actual spikes missed by the algorithm.

Figure \ref{fig:SyntheticMultichan} presents the measured spike detection accuracy across four noise levels in the synthetic dataset and across 1024 fabricated HRAM channels. The proposed IMC Ev-SPD maintained high detection accuracy across most fabricated channels in the low noise datasets (noise005 and noise01). 

However, performance degradation was observed on a small fraction of channels under lower SNR conditions, with average reductions of 3.27\% in noise015 and 12.5\% in noise02.
The reason for this can be traced to the effect of technology variations of $M_{cap}$ and $M_{res}$ in the bitcell. 
As an example, Monte Carlo simulations show that the jump in voltage on $M_{cap}$ per event may vary as shown in Fig. \ref{fig:MCandSyntheticResult}(a). \textcolor{black}{Further, to capture the effect of variability in threshold of the inverter when latching the result, we did a Monte Carlo simulation of the final voltage after thresholding. This result is shown in  Fig. \ref{fig:MCandSyntheticResult}(b) and Fig. \ref{fig:MCandSyntheticResult}(c) for the case of a positive and negative detection respectively. The results showed that the success rate for the SRAM latch to implement a negative (positive) detection is 97\% (98.3\%), validating the bitcell’s robustness to process variations. With the success rate statistics from MC simulation, we conducted an MC-informed software simulation by injecting bit flips in $A<i>$ (Eq. \ref{eq:A_i}). The results plotted in Fig.\ref{fig:MCandSyntheticResult}(d) show that the accuracy changed minimally (remained same for the low-noise datasets and dropped by only $\approx0.4\%$ on noise015 and noise20) due to the second level of filtering done by obtaining $S<i>$ (Eq. \ref{eq:S_i}) and applying a second precise digital threshold $Thr_2$ on that.
}

To compensate for variations, we introduce a channel-wise calibration scheme for the in-cell PMOS bias voltage ($V_{BP}$) in high-noise datasets. Channels with low sensitivity are configured to use a lower $V_{BP}$ to enhance firing, whereas those with a high FDR are assigned a higher $V_{BP}$ to suppress firing. The calibration is implemented with two configuration bits per channel to select between four $V_{BP}$ values. \textcolor{black}{These are not included in our current chip but we estimate the overhead for including them based on two SRAM cells as $\approx1$ $\mu m^2$ per channel , representing only $\approx0.26\%$ overhead.}

We constructed two canonical calibration stimuli: a spike waveform and a noise segment, respectively, from the noise02 dataset, each injected as AER events and presented at a period of 10 ms over a 1-second test for each stimulus (Fig. \ref{fig:calibration}). 
For the spike stimulus, we record the false-negative (FN) count; for the noise stimulus, we record the false-positive (FP) count. 
For a given channel, we select the bias value $v^{*}$ that minimizes the sum of FN and FP counts:
\begin{equation}
v^{*} = \arg\min \left( \mathrm{FN}_s(v) + \mathrm{FP}_n(v) \right)
\end{equation}
where $\mathrm{FN}_s(v)$ and $\mathrm{FP}_n(v)$ denote the FN and FP counts measured under the spike and noise stimuli, respectively. \textcolor{black}{Note that this calibration is based on the mismatch in filtering effect between channels and is ideally independent of the neural data being processed (e.g. Neuropixels dataset shown later). For higher precision, if the actual data has to be used for calibration, we can still extract and construct the spike stimuli and noise stimuli from the recorded events and use these for fine tuning the calibrated biases.}

Figure \ref{fig:MCandSyntheticResult} (d) summarizes the median spike detection metrics after calibration across four noise levels on the synthetic dataset. 
\textcolor{black}{
After calibration, our event-based IMC SPD macro achieved an accuracy ranging from 88\% to 99\% accuracy with a mean accuracy of 96.06\% across four noise levels, only 3.14\% lower than the software simulation implementing the EV-SPD algorithm in Section \ref{algorithm_sec} (that does not account for any hardware non-ideality). Further improvements in accuracy may be obtained by jointly optimizing the modulation threshold of the EBF with bitcell calibration to achieve the best SPD performance}.

\subsubsection{Neuropixel Recordings}
Unlike the synthetic dataset with ground truth spike times in its composition, the Neuropixel dataset are raw recordings without the spike time labels. To validate the proposed SPD, we labeled the spike time with a commonly used absolute threshold spike detection in \cite{quirogaUnsupervisedSpikeDetection2004} given by the empirical formula:
\begin{equation}
\operatorname{Thr_{abs}} = 4*\text{median}\left\{\frac{|x|}{0.6745}\right\}
\end{equation}

We fed 384-channel Neuropixel recordings to the macro and compared the spike times from the macro with the spike times from absolute thresholding. 
Fig. \ref{fig:NP}(b) shows the 384-channel spike detection accuracy using the ABS SPD as the ground truth. The SPD results from our IMC macro achieved a high mean similarity of 95.08\%. \textcolor{black}{Difference in SPD performance in this dataset compared to the earlier Synthetic one stem mainly from the huge variability of SNR in this dataset \cite{akhoundi1521024Channel000029mm22025}}.

To emulate the input to a primate reaching decoding task \cite{yikNeurobenchFrameworkBenchmarking2025}, we binned the detected spike times from both IMC Ev-SPD and ABS SPD in 4-ms windows to form the firing patterns in Fig. \ref{fig:NP}(a). The mean absolute error (MAE) between two firing patterns are only 0.05 calculated with the formula below:
\begin{equation}
\text{MAE} = \frac{1}{X \cdot Y} \sum_{i=1}^{X} \sum_{j=1}^{Y} |F_{\text{Ev}}(i,j) - F_{\text{ABS}}(i,j)|
\end{equation}

\begin{figure}[t]
    \centering
    \includegraphics[width=\columnwidth]{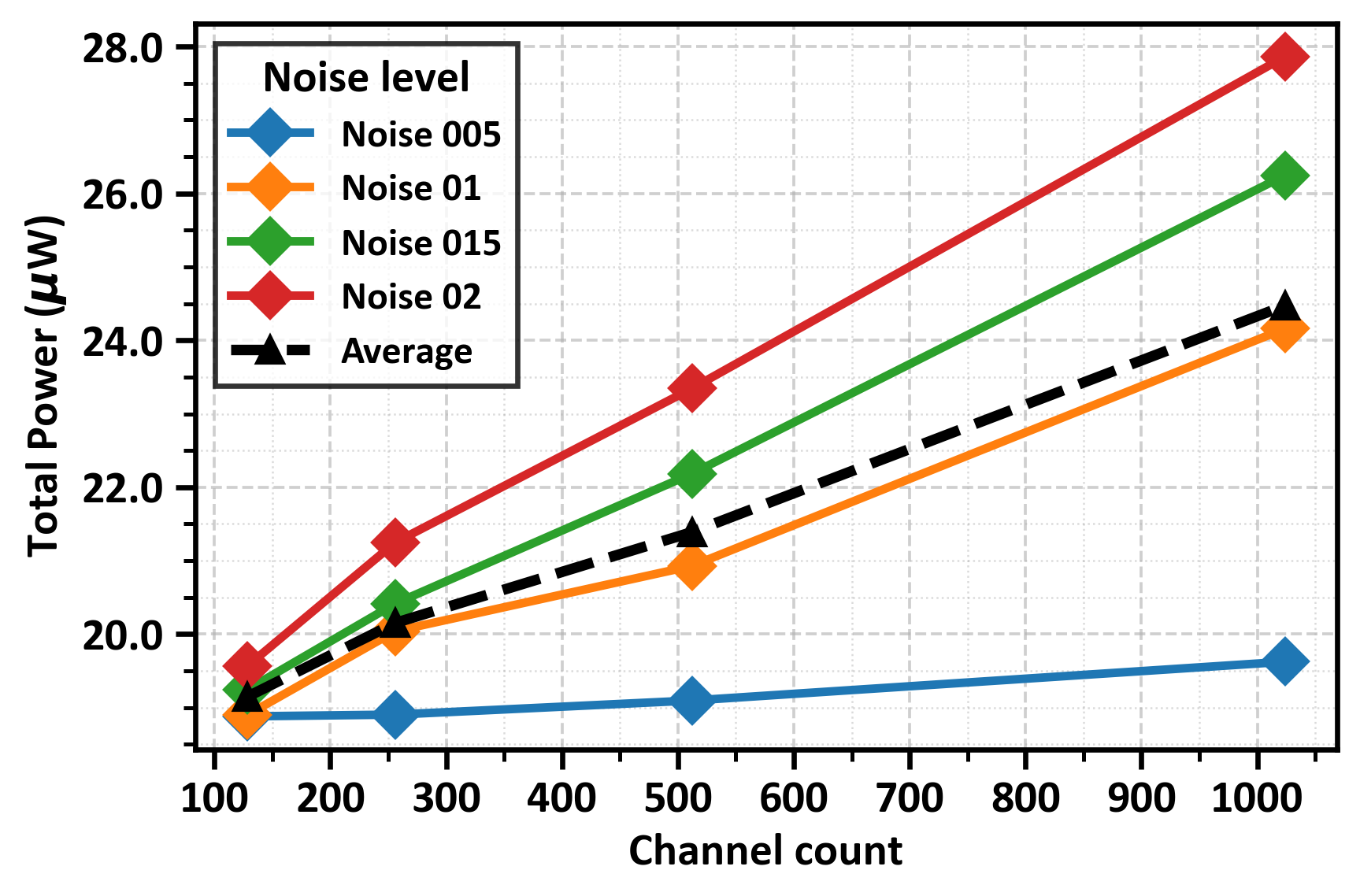}
    \caption{Total power consumption at different noise levels and for different number of channels}
    \label{fig:Power}
\end{figure}

\begin{figure}[t]
    \centering
    \includegraphics[width=\columnwidth]{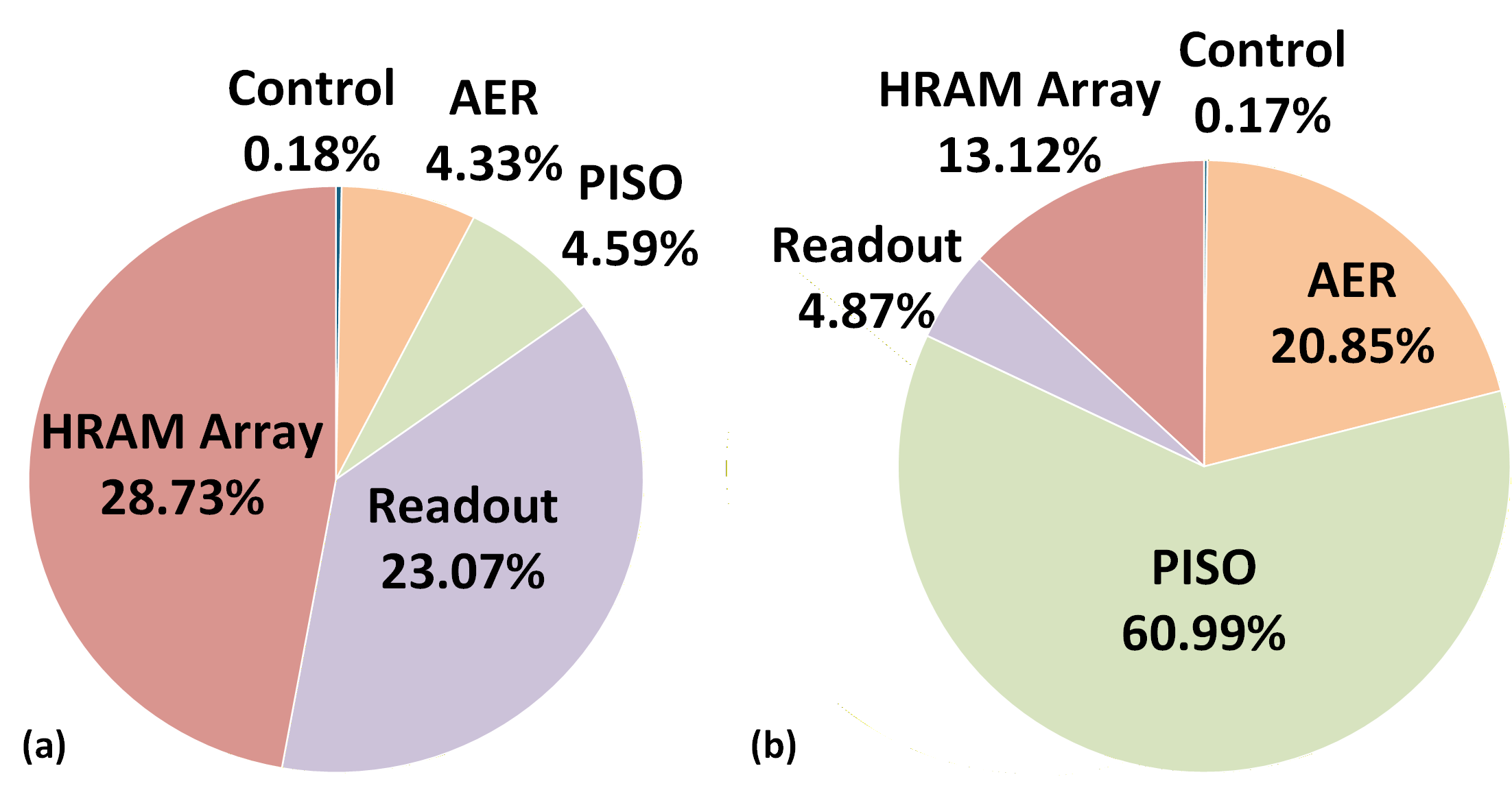}
    \caption{(a) Area breakdown (b) Energy breakdown of the IMC SPD macro.}
    \label{fig:Breakdown}
\end{figure}

\subsection{Discussion}
\label{sec:discussion}

At a 0.8-V supply, the 1024-channel IMC-SPD macro exhibits a measured static power of 3.04 $\mu\mathrm{W}$. 
\textcolor{black}{
We measured the total macro power at four noise levels, with event rates ranging from 2.1k to 6.8k events/s based on the synthetic dataset presented earlier, and with the events broadcast over 128, 256, 512, and 1024 channels (Fig. \ref{fig:Power}).
For 1024-channel operation over the synthetic dataset, the measured power averaged across four noise levels is 24.47 $\mu\mathrm{W}$;
}
the breakdown is shown in Fig. \ref{fig:Breakdown}(b). 
The HRAM array accounts for only 3.21 $\mu\mathrm{W}$ on average,
\textcolor{black}{ while the total power is dominated by the PISO circuitry.
The high power consumption of the PISO block in testing is partially due to the charging /discharging of large capacitance related to the I/O pad and PCB traces--this will reduce at least by an order of magnitude for a fully on-chip system.
For future system-level full integration of HRAM SPD in event-based BMI implementations, the readout overhead could be reduced further by replacing the PISO module with a digital comparator cycling through the ripple counter values and comparing them with a global threshold $Thr_2$ to produce an output for every channel per $T_s$ detection cycle. Moreover, the in-pixel integration of the HRAM spike detector with the event-based frontend could compress the AER event rate to neuron firing rate and reduce the system-level communication overhead of the event-based sensing array. Also, our current design does not feature channel-wise self-adaptive thresholds. This can be handled using EBF with adaptive thresholds or by tuning the in-bitcell PMOS bias in our proposed design by background calibration. These will be handled in our future designs.}

In Table. \ref{tab:comparison} we compared the proposed IMC SPD with recent spike detector implementations. 
\textcolor{black}{
Our work supported the highest neural channel count and achieved the lowest per-channel power of 23.9 nW/Ch and a compact per-channel area of 375 $\mu m^2$/Ch.}
Compared with the digital ASIC simulation for Nyquist frontends presented in \cite{zhangCalibrationFreeHardwareEfficientNeural2023}, our IMC spike detector achieved a significant reduction in memory array size, avoided frequent memory access, and brought about lower computation overhead. 
\textcolor{black}{In comparison with the digital valid counter and NEO implementation in \cite{akhoundi1521024Channel000029mm22025}, our in-memory computing approach achieves power savings by reducing the input buffering overhead.}
Moreover, the proposed event-based macro is compatible with compressive EBFs, which could improve scalability and enable low-power spiking neural network decoders \cite{biyanCombiningSNNsFiltering2025}.
Relative to the integrate-and-fire digital spike detector implementation in \cite{chenNeuronInspired00032mm2138mW2024}, which is also designed for compressive EBFs, our IMC architecture enables more effective event processing and supports a higher channel count. 
\textcolor{black}{We have also compared this work with other eDRAM-based in-memory computing bitcells in Table \ref{tab:edram_comparison}. Our eDRAM design enables a low write energy of 0.5 fJ per event, and our shared detection line has a low read energy of 4.4 fJ per bit, achieving a much higher energy efficiency.}
The sparse computing nature of our IMC macro significantly improved the energy efficiency of spike detection.
The compact size of proposed IMC bitcells has also enabled the in-pixel integration with compressive EBFs, which may significantly reduce the number of AER events from the frontend and further improve system-level scalability.

\begin{table}[!t]
\huge
\centering
\caption{\textcolor{black}{Comparison with prior eDRAM-based IMC Bitcells}}
\vspace{-20pt}
\resizebox{\columnwidth}{!}{%
{\huge
\setlength{\arrayrulewidth}{0.6pt}
\renewcommand{\arraystretch}{1.3}  
\begin{threeparttable}
\label{tab:edram_comparison}
\begin{tabular}{lccccc}
\toprule
 & \textbf{This Work} & \textbf{JSSC23\cite{zhang9151220TOPS97613012023}} & \textbf{JSSC24\cite{xieEDRAMCIMReconfigurableCharge2024}} & \textbf{TCAS21\cite{yuLogicCompatibleEDRAMComputeInMemory2021}} & \textbf{ISSCC21\cite{chen15365nm3T2021}} \\
\midrule
\makecell{Technology\\ (nm) }
& 65 & 65 & 65 & 65 & 65\\

\makecell{Cell Type }& \makecell{eDRAM\\+SRAM} & \makecell{eDRAM\\+SRAM} & eDRAM & eDRAM & eDRAM \\

\makecell{Cell Detail }& 9T1C& 10T1C& 1T1C & 4T2C& 3T1C\\

\makecell{Application} & \makecell{Spike\\Detection} & \makecell{DVS\\Processing} & MAC & MAC & MAC\\

\makecell{Supply\\Voltage (V)} & 0.8 & 0.7--1.2 & 1--1.2 & 0.5 & 1\\

\makecell{Write Energy} & 0.5 fJ/event & -- & -- & -- & -- \\

\makecell{Read Energy} & 4.4 fJ/bit & -- & 30.6 fJ/MAC & 26.2 fJ/OP & --\\

\makecell{Bitcell Area\\($\mu$m\textsuperscript{2})} & 9.7 & 4 & 22.08 & 1.345 & ~1.5\\

\makecell{Macro Size\\(bit)} & 8k & 10k & 16k & 16k & 8k\\
\bottomrule
\end{tabular}
\end{threeparttable}
}
}
\end{table}

\section{Conclusion}

In this work, we implemented a novel area and power-efficient event-based IMC SPD macro for compressive event-based EBFs. It uses a combination of SRAM and embedded DRAM to achieve event filtering and thresholding in the memory bitcell. Further robustness is achieved by in-memory averaging of the detection result across multiple time bins.
The chip achieved a high per-channel SPD energy efficiency of 23.9 nW per channel and a high area efficiency of 375 $\mu m^2$ per channel. 
The proposed IMC SPD demonstrated a high accuracy of 96.06\% on synthetic dataset, and 95.08\% similarity and 0.05 firing pattern MAE on Neuropixel recordings.
Our work presents an AER-compatible in-memory computing spike detection scheme for next-generation high-channel-count iBMIs, achieving ultra-low power and area overhead.

\section*{Acknowledgments}
This work was supported by the Research Grants Council of the HK SAR, China (Project No. CityU 11200922 and HKU C7003-24Y).

{\appendix[\textcolor{black}{Ev-SPD Simulation on meaREC Datasets}]
\textcolor{black}{To validate the adaptability of the proposed EV-SPD algorithm at different SNRs, we incorporated the meaREC simulator \cite{buccinoMEArecFastCustomizable2021} widely used in recent spike sorting works\cite{akhoundi1521024Channel000029mm22025} to simulate a neuron-electrode model with a firing rate of 20 Hz. We generated twenty 6-second synthetic extracellular recordings with SNRs varying from 4 dB to 76dB.} 

\textcolor{black}{Then we applied the Ev-SPD algorithm to these recordings and performed a parameter sweep evaluation. The resulting accuracy heatmap is shown in Fig. \ref{fig:Append} (a). The results indicate that EV-SPD maintains high accuracy across a broad set of hyperparameter combinations.
In Fig. \ref{fig:Append} (b), we compared the accuracy of Ev-SPD and NEO with respect to the SNR of recordings. The results show that our proposed event-based algorithm achieved a similar accuracy performance with NEO across a wide range of SNRs.
We also conducted a sensitivity analysis in which the window $T_s$ was varied by $\pm20\% $. The results show that the algorithm remains stable for variations in $T_s$.}

\begin{figure}[htp]
    \centering
    \includegraphics[width=\columnwidth]{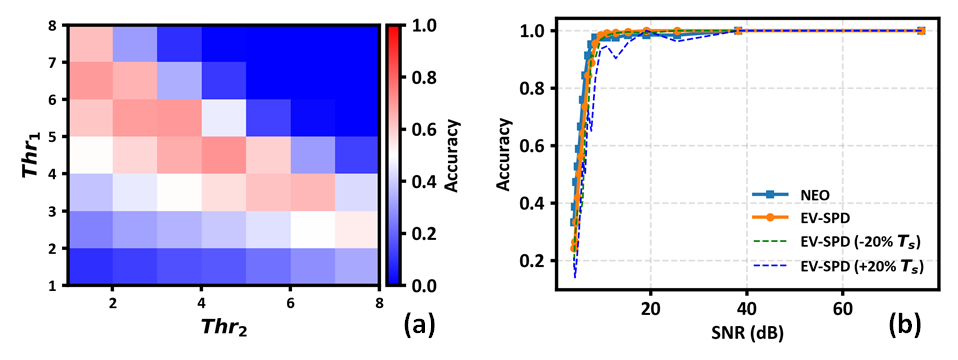}
    \caption{\textcolor{black}{(a) Ev-SPD algorithm accuracy on the meaREC recordings at different parameter settings. (b) Comparison between Ev-SPD and NEO algorithm accuracy at different recording SNRs.}}
    \label{fig:Append}
\end{figure}


%

\bibliographystyle{IEEEtran} 
\bibliography{refs.bib}


\section{Biography Section} 

\vspace{-33pt}
\begin{IEEEbiography}[{\includegraphics[width=1in,height=1.25in,clip,keepaspectratio]{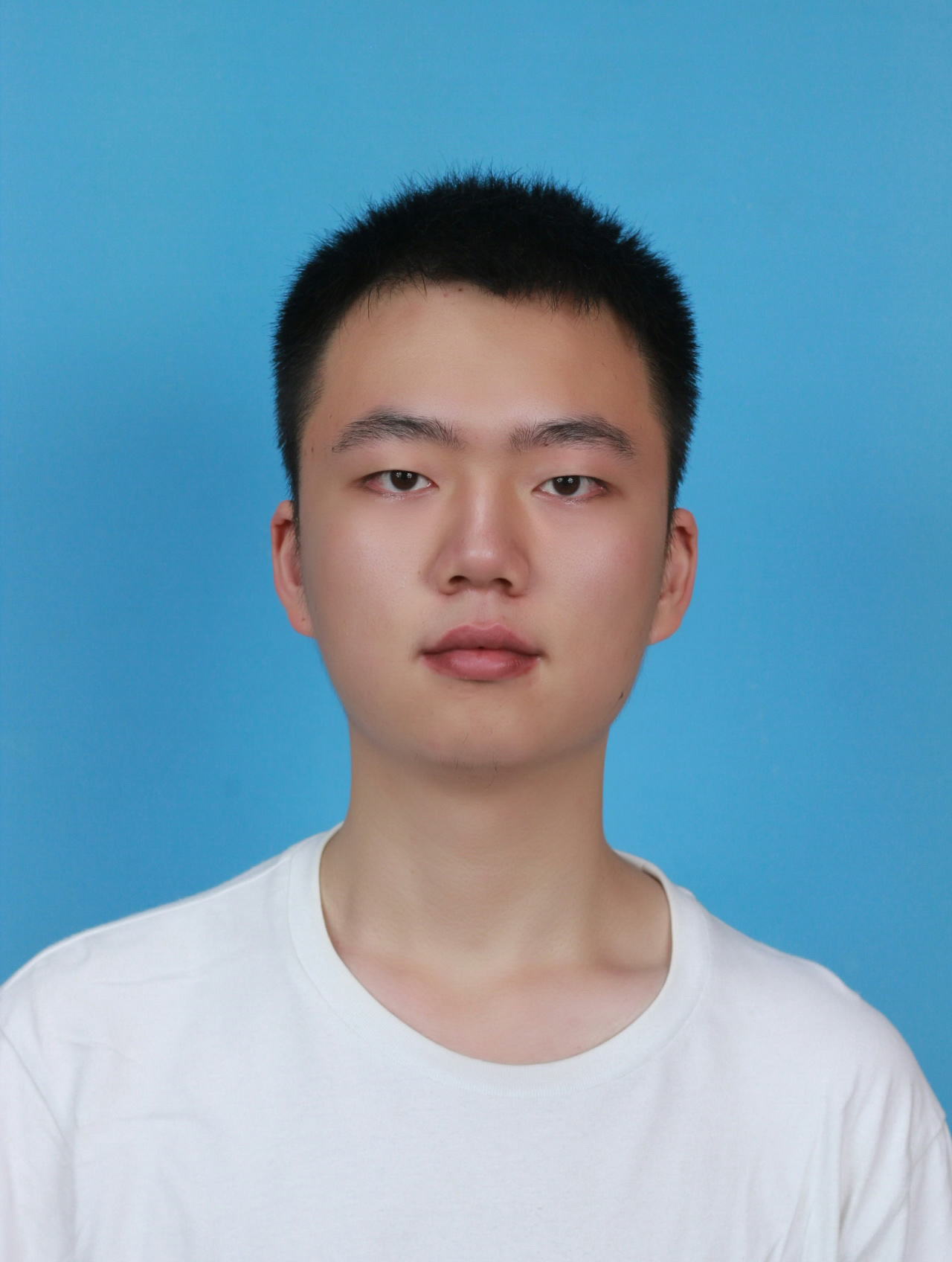}}]{Ye Ke}
received joint B.Eng. (Hons.) degree in Electronic and Information Engineering from University of Electronic Science and Technology of China (UESTC) and University of Glasgow. 
He is currently pursuing a PhD degree at the City University of Hong Kong. His research interests span brain-machine interfaces, in-memory computing, and spiking neural networks.
\end{IEEEbiography}

\vspace{-33pt}
\begin{IEEEbiography}[{\includegraphics[width=1in,height=1.25in,clip,keepaspectratio]{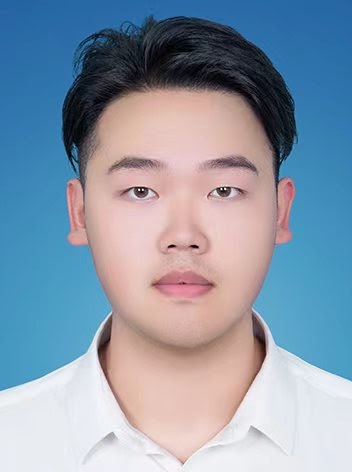}}]{Zhengnan Fu}
received his bachelors from Central South University (CSU) and M.Sc from University of Chinese Academy of Sciences (UCAS), Shanghai Institute of Microsystem and Information Technology (SIMIT). He is currently a PhD student in BRAIN lab at CityU, Hong Kong working on brain-machine interface circuits, neuromorphic sensor systems front-end, and in-memory computing circuit.
\end{IEEEbiography}

\vspace{-33pt}
\begin{IEEEbiography}[{\includegraphics[width=1in,height=1.25in,clip,keepaspectratio]{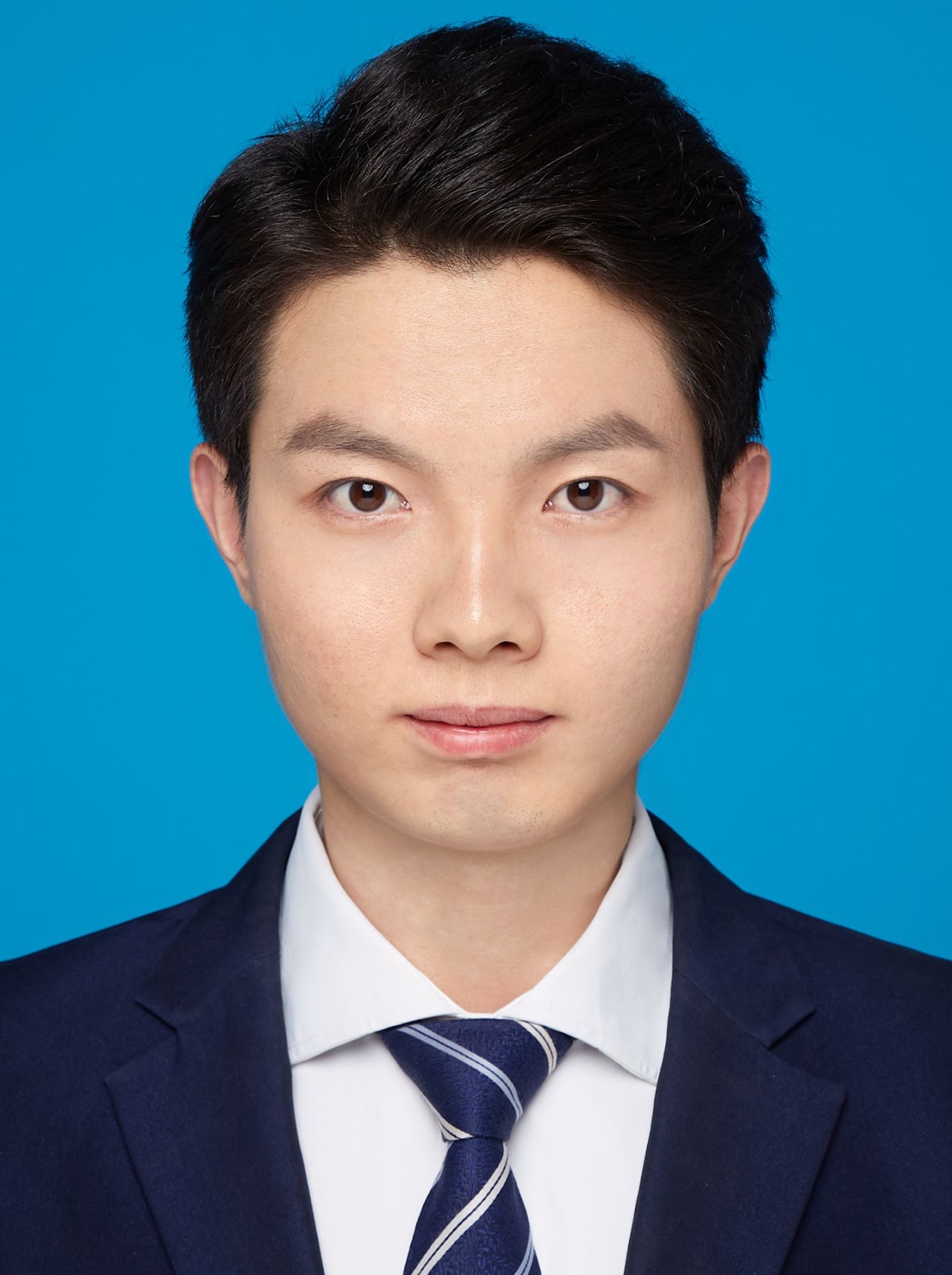}}]{Junyi Yang}
 received his Bachelors from China University of Mining and Technology and M.Sc from Beihang University with the excellent graduate award of Beijing. He is currently working in BRAIN lab on In-memory computing using volatile and non-volatile memories for neuromorphic systems.
\end{IEEEbiography}

\vspace{-33pt}
\begin{IEEEbiography}[{\includegraphics[width=1in,height=1.25in,clip,keepaspectratio]{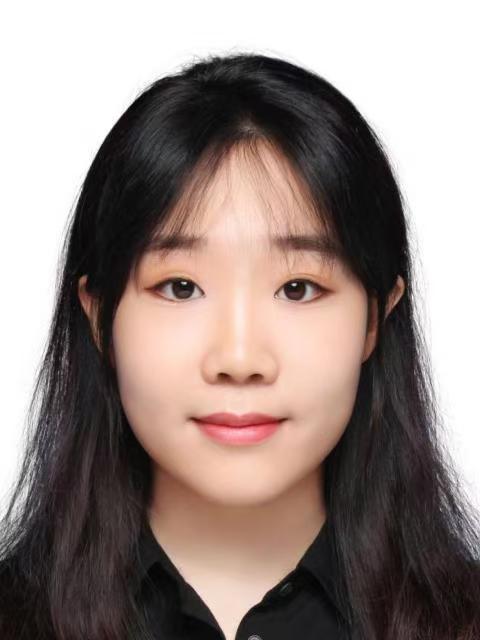}}]{Hongyang Shang}
received her B.S. degree in electronic science and technology from Nankai University, and M.S. degree in electronics from Nanyang Technology University, respectively. After that, she worked as a research assistant in Fudan University. She is currently a PhD student in the Department of Electrical Engineering at City University of Hong Kong. Her research interests include bio-inspired neuromorphic circuits, event-based camera and computing-in-memory.

\end{IEEEbiography}

\vspace{-33pt}
\begin{IEEEbiography}[{\includegraphics[width=1in,height=1.25in,clip,keepaspectratio]{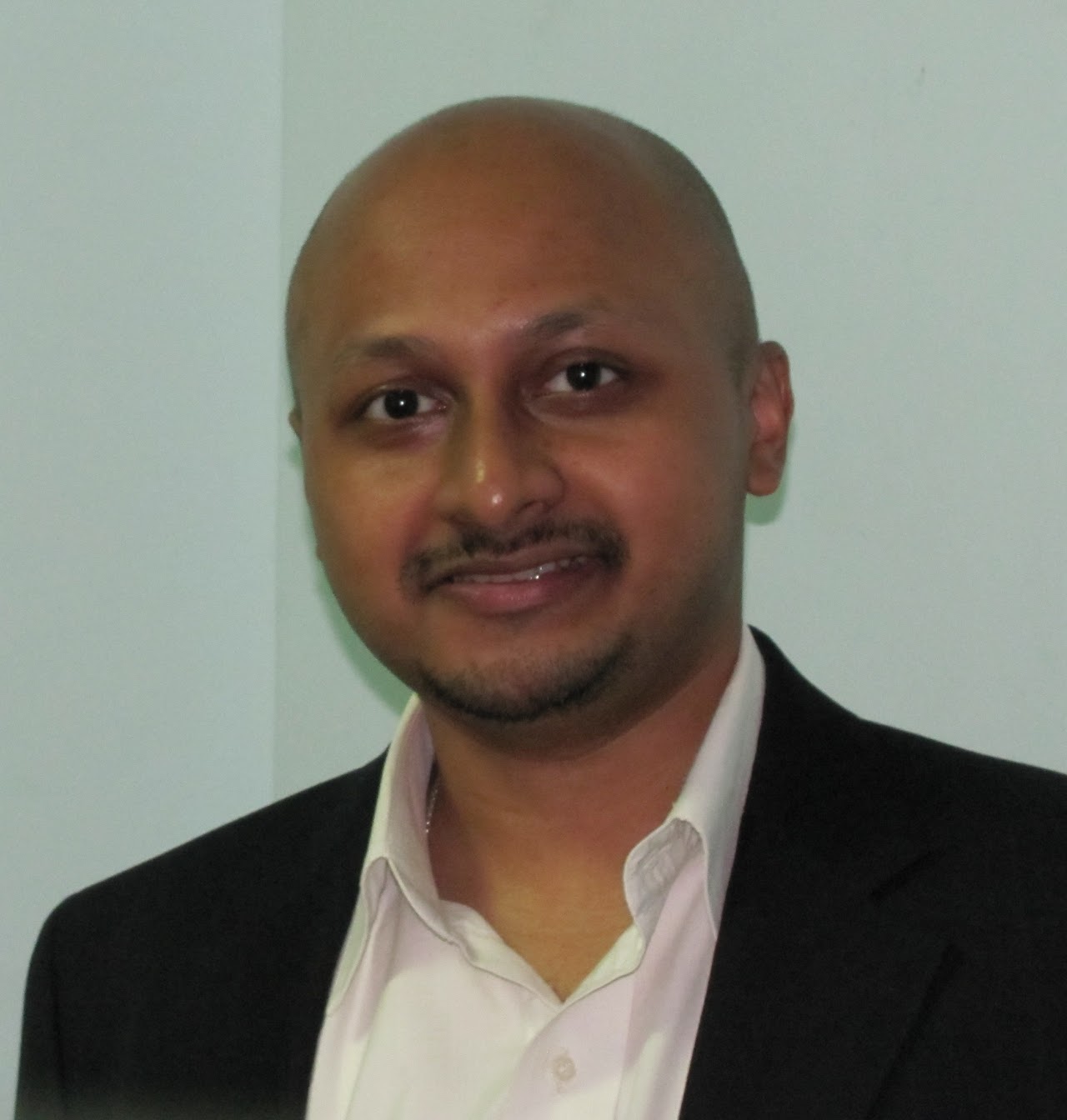}}]{Arindam Basu}
 (Senior Member, IEEE) received the B.Tech. and M.Tech. degrees in electronics and electrical communication engineering from Indian Institute of Technology Kharagpur, in 2005, and the M.S. degree in mathematics and the Ph.D. degree in electrical engineering from Georgia Institute of Technology, Atlanta, in 2009 and 2010, respectively. He was a tenured Associate Professor with Nanyang Technological University. He is currently a Professor with the City University of Hong Kong, Department of Electrical Engineering. Dr. Basu is a Technical Committee Member of the IEEE CAS Societies of Biomedical Circuits and Systems, Sensory Systems and Neural Systems and Applications (Past Chair). He received the Prime Minister of India Gold Medal from IIT Kharagpur in 2005. He is currently the Associate Editor-in-Chief of IEEE TRANSACTIONS ON BIOMEDICAL CIRCUITS AND SYSTEMS and an Associate Editor of IEEE SENSORS JOURNAL, Frontiers in Neuroscience, and Neuromorphic Computing and Engineering (IOP). He has served as an IEEE CAS Distinguished Lecturer (2016–2017). He received the Best Student Paper Award at Ultrasonics Symposium in 2006, Best Live Demonstration at ISCAS 2010, and a Finalist position in the Best Student Paper Contest at ISCAS 2008. He was awarded MIT Technology Review’s TR35 Asia–Pacific Award in 2012 and inducted into Georgia Tech Alumni Association’s 40 under 40 class of 2022.
\end{IEEEbiography}

\vspace{11pt}

\vfill

\end{document}